\documentclass[
preprint,           % 12pt, single-column
superscriptaddress, % authors with affiliations via superscripts
amsmath,            % add AMS-Latex features
amssymb,            % add extra AMS symbols, including amsfonts
aps,                % aps or aip
pr,                 % prl, pra, prb, prc, prd, pre, prstab
longbibliography,   % show article titles in the bibliography
floatfix,           % process floats as early as possible
nofootinbib,        % show footnote at the text
]{revtex4-1}

\usepackage{tensor}     % manipulate tensors
\usepackage{graphicx}   % include figures
\usepackage[
colorlinks=true,        % color link
citecolor=blue,         % cite color
linkcolor=blue,         % link color
urlcolor=blue          % url color
]{hyperref}             % create hyperlinks
\usepackage{bm}         % \bm{<text>} Bold math symbols
\usepackage{xcolor}     % textcolor
\usepackage{multirow}
\newcommand{\newc}{\newcommand*}
\newc{\figurewidth}{3.2in}
\newc{\la}{\lambda}
\newc{\La}{\Lambda}
\newc{\io}{\iota}
\newc{\Msun}{M_\odot}
\newc{\lvc}{LIGO/Virgo}
\newc{\PBH}{\mathrm{PBH}}
\newc{\ABH}{\mathrm{ABH}}
\newc{\fpbh}{f_\mathrm{PBH}}

\def\({\left(}
\def\){\right)}
\def\[{\left[}
\def\]{\right]}

\def\del{((\Delta \sigma_q)^{2}+(\Delta \sigma_g)^{2})}

\def\ee{\begin{equation}}
\def\q{\end{equation}}
\def\m{\begin{eqnarray}}
\def\n{\end{eqnarray}}
\def\a{\begin{aligned}}
\def\b{\end{aligned}}

\newc{\red}[1]{\textcolor{red}{#1}}
\newc{\yellow}[1]{\textcolor{yellow}{#1}}
\newc{\green}[1]{\textcolor{green}{#1}}
\newc{\blue}[1]{\textcolor{blue}{#1}}
\begin{document}
\title{Gravitational and electromagnetic radiation from binary black holes with electric and magnetic charges: Hyperbolic orbits on a cone}
%%%%%%%%%%%%%%%%%%%%%%%%%%%%%%% author  %%%%%%%%%%%%%%%%%%%%%%%%%%%%%
\author{Zu-Cheng Chen}
\email{zucheng.chen@bnu.edu.cn}
\affiliation{Department of Astronomy, Beijing Normal University, Beijing 100875, China}
\affiliation{Advanced Institute of Natural Sciences, Beijing Normal University, Zhuhai 519087, China}
\affiliation{Department of Physics and Synergistic Innovation Center for Quantum Effects and Applications, Hunan Normal University, Changsha, Hunan 410081, China}

\author{Sang Pyo Kim}
\email{sangkim@kunsan.ac.kr}
\affiliation{Department of Physics, Kunsan National University, Kunsan 54150, Korea}

\author{Lang Liu}
\email{Corresponding author: liulang@bnu.edu.cn}
\affiliation{Department of Astronomy, Beijing Normal University, Beijing 100875, China}
\affiliation{Advanced Institute of Natural Sciences, Beijing Normal University, Zhuhai 519087, China}

%%%%%%%%%%%%%%%%%%%%%%%%%%%%%%%%%%%%%%%%%%%%%%%%%%%%%%%%%%%%%%%%%%%%%%
\date{\today}
%%%%%%%%%%%%%%%%%%%%%%%%%%%%%%%%%%%%%%%%%%%%%%%%%%%%%%%%%%%%%%%%%%%%%%
\begin{abstract}
We derive the hyperbolic orbit of binary black holes with electric and magnetic charges. In the low-velocity and weak-field regime, by using the Newtonian method, we calculate the total emission rate of energy due to gravitational and electromagnetic radiation from binary black holes with electric and magnetic charges in hyperbolic orbits. Moreover, we develop a formalism to derive the merger rate of binary black holes with electric and magnetic charges from the two-body dynamical capture. We apply the formalism to investigate the effects of the charges on the merger rate for the near-extremal case and find that the effects cannot be ignored.  
\end{abstract}
%\pacs{???}
\maketitle
%%%%%%%%%%%%%%%%%%%%%%%%%%%%%%%%%%%%%%%%%%%%%%%%%%%%%%%%%%%%%%%%%%%%%%
%%%%%%%%%%%%%%%%%%%%%%%%%%%%%%%%%%%%%%%%%%%%%%%%%%%%%%%%%%%%%%%%%%%%%%
\section{Introduction}\label{intro}
The first successful measurement of gravitational-wave (GW) signal \cite{Abbott:2016blz} from a compact binary coalescence by LIGO has marked the dawn of multi-messenger astronomy and opened a new window to probe the Universe \cite{LIGOScientific:2018mvr,Abbott:2020niy,LIGOScientific:2021djp}. GWs are also a powerful tool for testing gravity theory in the strong-field regime. So far, the merger events from  LIGO-Virgo-KAGRA (LVK) collaboration can all be well described by general relativity (GR) \cite{LIGOScientific:2019fpa,LIGOScientific:2020tif,LIGOScientific:2021sio}.

The no-hair theorem of black holes (BHs) in GR states that a general relativistic BH is completely described by four physical parameters: mass, spin, electric charge, and magnetic charge. If magnetic charges exist in the universe, they can provide a new unexplored window into fundamental physics in the Standard Model of particle physics. Although no evidence of magnetic charges has been found in the laboratory until now \cite{Staelens:2019gzt,Kobayashi:2021des}, GWs provide a completely different way to test magnetic charges. Magnetically charged BHs have attracted much attention not only in theoretical study but also in recent astronomical observations \cite{Maldacena:2020skw, Bai:2020spd, Liu:2020vsy, Ghosh:2020tdu, Liu:2020bag}. For instance, Ref.~\cite{Maldacena:2020skw} discusses the spectacular properties of magnetically charged BHs, showing that the magnetic field near the horizon of the magnetically charged BH can be strong enough to restore the electroweak symmetry. The astrophysical signatures for magnetically charged BHs also have been studied in Ref.~\cite{Ghosh:2020tdu}.

Compared with Schwarzschild BHs, charged BHs emit both gravitational and electromagnetic radiation and have rich phenomena. Recently, there has been an increasing interest in charged BHs; see Refs.~\cite{Liu:2020cds,Maldacena:2020skw,Bai:2020spd,Liu:2020vsy,Ghosh:2020tdu,Liu:2020bag,Zilhao:2012gp,Zilhao:2013nda,Liebling:2016orx,Toshmatov:2018tyo,Bai:2019zcd,Allahyari:2019jqz,Christiansen:2020pnv,Wang_2021,Bozzola:2020mjx,Kim:2020bhg,Cardoso:2020nst, McInnes:2020gxx,Bai:2020ezy,Diamond:2021scl,Bozzola:2021elc,McInnes:2021frb,Kritos:2021nsf,Hou:2021suj,Benavides-Gallego:2021the,Diamond:2021dth,Ackerman:2008kmp,Feng:2009mn,Foot:2014uba,Foot:2014osa,Moffat:2005si,Cardoso:2016olt,Cardoso:2020iji,Liu:2022cuj,Liu:2022wtq,Zhang:2022hbt,Pina:2022dye, Zi:2022hcc,Benavides-Gallego:2022dpn,Estes:2022buj} and references therein. In the previous papers \cite{Liu:2020vsy, Liu:2020bag}, we studied the case of binary BHs (BBHs) with electric and magnetic charges in circular and elliptical orbits on a cone. On the one hand, using the Newtonian approximation with radiation reactions, we calculate the total emission rate of energy and angular momentum due to gravitational and electromagnetic radiation. In the case of circular orbits, we show that electric and magnetic charges could significantly suppress the merger times of the dyonic binary system.
On the other hand, when considering elliptical orbits, we show that the emission rate of energy and angular momentum due to gravitational and electromagnetic radiation have the same dependence on the conic angle for different cases. {Not all BBHs are bounded systems and those from encounters of black holes could have positive energy. Therefore, it is important and meaningful to derive the orbit of BBHs with electric and magnetic charges for the unbounded case (i.e.~$E>0$) and explore their characteristic features.}

{In this paper, we extend our previous analyses to the unbounded case and derive the hyperbolic orbit of BBHs with electric and magnetic charges.}
%In this paper, we extend our previous analyses to the hyperbolic case. 
In the Universe, the two-body dynamical capture is an absolutely common and effective way to form BBH systems. We also derive the merger rate of BBHs with electric and magnetic charges from the two-body dynamical capture.
The paper is structured as follows. In section \ref{II}, we derive the hyperbolic orbit of BBHs with electric and magnetic charges. In the low-velocity and weak-field regime, by using a Newtonian method, we calculate the total emission rate of energy due to gravitational and electromagnetic radiation from BBHs with electric and magnetic charges in hyperbolic orbits. In section \ref{III}, we develop a formalism to derive the merger rate of BBHs with electric and magnetic charges from dynamical capture via gravitational and electromagnetic radiation. In section \ref{IV}, we apply the formalism to find the effects of the charges on the merger rate for the near-extremal case and discover that the effects cannot be ignored. Finally, section \ref{Con} is devoted to the conclusion and discussion. Throughout this paper, we set $G=c =4 \pi \varepsilon_{0} = \frac{\mu_0}{4\pi}=1$ unless otherwise specified.

\section{\label{II}Gravitational and electromagnetic radiation}

In this section, we focus on the case that the distance of the dyonic BH binary is much larger than their event horizons. In such a case, the metric is approximately the Minkowski metric. Therefore, it is a good approximation that the dyonic BH binary is described by two massive point-like objects with electric and magnetic charges in the Minkowski spacetime. This approximation has also been employed in recent works \cite{Bai:2019zcd, Liu:2020cds, Ghosh:2020tdu} that examine charged binary black holes. Recent numerical-relativity simulations validate this approximation when the separation distance between the black hole binary is significantly larger than their event horizons \cite{Bozzola:2021elc}. 
%Note that although the orbital motion is hyperbolic-like, the gravitational and electromagnetic radiation is a relativistic outcome, similar to the post-Newtonian gravitational radiation from a massive binary~\cite{Peters:1963ux,Peters:1964zz} and the synchrotron radiation from electric and magnetic charges in a hyperbolic orbit~\cite{1975ctf..book.....L}.}
%We adopt the weak-field approximation by assuming the distance of two dyonic BHs is much larger than their event horizons. We find that this condition could be safely satisfied for the hyperbolic case. 
Before we calculate the total emission rate of energy due to gravitational and electromagnetic radiation from BBHs with electric and magnetic charges, we need to know the hyperbolic orbit. In the following subsection, we will derive the hyperbolic orbit of BBHs with electric and magnetic charges.

\subsection{Hyperbolic orbits of BBHs with electric and magnetic charges without radiation}
\label{sub-1}
Here, we consider the hyperbolic encounter of two BHs with mass, electric and magnetic charges ($m_1$, $q_1$, $g_1$) and ($m_2$, $q_2$, $g_2$).  According to Refs.~\cite{Liu:2020vsy,Liu:2020bag}, choosing the center of mass system at the origin and considering the Lorentz force and gravitational force, the equation of motion is
\m
\mu \ddot{R}^{i}=C \frac{R^{i}}{R^{3}}-D \epsilon_{ j k}^{i} \frac{R^{j}}{R^{3}} v^{k},
\n
where $R$ is the distance between two dyonic BHs, $C=-m_1 m_2+q_{1} q_{2}+g_{1} g_{2}$, $D=q_{2} g_{1}-g_{2} q_{1}$, $v^{i}=\dot{R}^{i}=d R^i/d t$, and $\mu=\frac{m_1m_2}{m_1+m_2}$ is the reduced mass. Notice that $q_1^2+g_1^2 \leq m_1^2$ and $q_2^2+g_2^2 \leq m_2^2$, so $C \leq 0$. Following Refs.~\cite{Liu:2020vsy, Liu:2020bag}, the generalized angular momentum of the binary system $\bm{L}$ is the Laplace-Runge-Lenz vector defined by $\bm{L}\equiv \bm{\tilde{L}}-D \hat{\bm{r}}$, where $\bm{\tilde{L}}\equiv \mu \bm{R} \times \bm{v}$ is the {orbital} angular momentum of binary system and $\hat{\bm{r}}$ is the unit vector along $\bm{R}$. It should be noted that one BH with an electric charge and the other BH with a magnetic charge also give a non-zero $D$, and therefore the orbits occur on the cone, as will be explained below.

\begin{figure}[htpb]
  \centering   \includegraphics[width=0.48\textwidth]{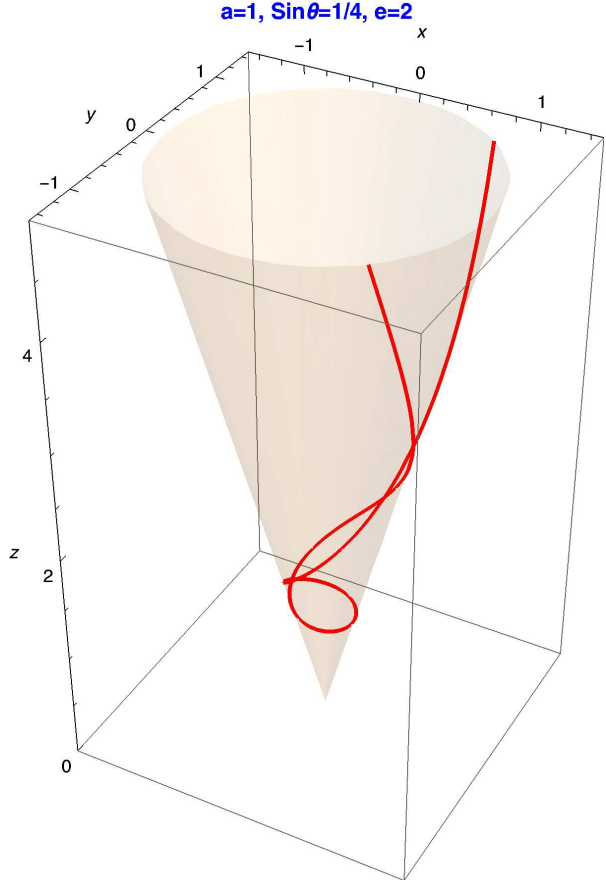}
    \caption{A conic-shaped orbit of the binary that is confined to the surface of a cone by choosing the parameters $a=1$, $e=2$, and $\sin \theta=1/4$.}
    \label{fig:R}
\end{figure}

Choosing the $z$-axis along the generalized angular momentum $\bm{L}$, the conserved module of the generalized angular momentum and energy are given by
\m
\label{EL}
L=\frac{\tilde{L}}{\sin \theta}=\mu R^2 \dot{\phi}, \quad
E=\frac{1}{2} \mu \dot{R}^{2}+\frac{\tilde{L}^{2}}{2 \mu R^{2}}+\frac{C}{R},
\n
where $\theta$ is a constant determined by $\cos \theta=\frac{|D|}{L}$. Throughout this paper, we only consider $\theta  \in (0,{\pi/2}]$ for simplify. For $\theta  \in [\pi/2,{\pi})$, we can refine $\boldsymbol{R}^{\prime}=-\boldsymbol{R}$ to make $\theta  \in (0,{\pi/2}]$. From Eq.~\eqref{EL}, eliminating the parameter $t$, we can get
\m
\frac{\dot{\phi}}{\dot{R}}=\frac{d \phi}{d R}=\left(\frac{2 \mu E}{L^{2}} R^{4}-\frac{2 \mu C}{L^{2}} R^{3}-\sin ^{2} \theta R^{2}\right)^{-\frac{1}{2}}.
\n
Adjusting $x=1/R$ and using the integral $\int \frac{d x}{\sqrt{\alpha+\beta x+\gamma x^{2}}}=\frac{1}{\sqrt{-\gamma}} \arccos \left(-\frac{\beta+2 \gamma x}{\sqrt{\beta^{2}-4 \alpha \gamma}}\right)$, we can get one of the solutions as $R=\frac{\frac{\tilde{L}^{2}}{\mu|C|}}{1+\sqrt{1+\frac{2 \tilde{L}^{2}}{\mu C^{2}} E} \cos ((\phi-\phi_0 )\sin \theta)}$, which is consistent with Refs.~\cite{Liu:2020vsy,Liu:2020bag}. Notice the sum, $\arccos (x)+\arccos (-x)$, is a constant, we can get the other branch of solution, $R=\frac{\frac{\tilde{L}^{2}}{\mu|C|}}{1-\sqrt{1+\frac{2 \tilde{L}^{2}}{\mu C^{2}} E} \cos ((\phi-\phi_0 )\sin \theta)}$. 
For the hyperbolic case in which $E > 0$, we choose $\phi_0=0$ for simplicity.
The two solutions result in an identical energy emission rate, so we only consider the second branch of the solution. Therefore, the orbit is explicitly given by
\m
\label{R}
\boldsymbol{R}=\frac{a\left(e^{2}-1\right)}{1-e \cos (\phi \sin \theta)}\left(\begin{array}{c}
	\sin \theta \cos \phi \\
	\sin \theta \sin \phi \\
	\cos \theta
\end{array}\right),
\n
where $a$ and $e$ can be interpreted as the semimajor axis and eccentricity. They are defined by
\m
a \equiv\left|\frac{C}{2 E}\right|= -\frac{C}{2E}, \quad e \equiv \left(1+\frac{2E \tilde{L}^{2}}{\mu C^2}\right)^{1/2}.
\n
Since $R>0$, we can derive the range as $\phi \in \left(\phi_1, \frac{2 \pi }{\sin \theta }-\phi_1 \right)$, where $\phi_1=\frac{\arccos \left(e^{-1}\right)}{\sin \theta}$.
Noting that $\theta$ is a constant determined by the initial condition, we can interpret Eq.~\eqref{R} as conic-shaped orbits of the binary, which is confined to the surface of a cone with half-aperture angle $\theta$. The orbits are shown in Fig.~\ref{fig:R} by choosing $a=1$, $e=2$, and $\sin \theta=1/4$.
Now we have the hyperbolic orbit, and we will calculate the total emission rates of energy due to gravitational and electromagnetic radiation in the next subsection.

\subsection{Gravitational and electromagnetic radiation from BBHs with electric and magnetic charges}
\label{sub-2}
In this subsection, we will focus on gravitational and electromagnetic radiation. Let us start by considering gravitational radiation. According to Ref.~\cite{Peters:1963ux}, the radiated power of GWs due to gravitational quadrupole radiation, $\frac{dE_{\mathrm{GW}}^{\mathrm{quad}}}{dt}$, is expressed as
\ee
\label{PGW}
\frac{dE_{\mathrm{GW}}^{\mathrm{quad}}}{dt}=-\frac{1}{5}\left(\dddot{Q}_{ij}\dddot{Q}_{ij}\right)=-\frac{1}{5 }\left(\dddot{M}_{i j} \dddot{M}_{i j}-\frac{1}{3}\left(\dddot{M}_{k k}\right)^{2}\right),
\q
where $M_{ij}$ is the second mass moment, and $Q_{ij} \equiv M_{ij}-\frac{1}{3} \delta_{ij} M_{kk}$ is traceless second mass moment. In our reference frame where $\bm{L}$ is along $z$ axis, the second mass moment $M_{ij}$ takes the $3 \times 3$ matrix form as
\ee
\label{Mij}
M^{ij}=\mu R_i R_j=\mu R^{2}\left(
\begin{array}{ccc}
	\sin^2 \theta \cos^2 \phi & \sin^2 \theta \cos \phi \sin \phi & \sin \theta \cos \theta \cos \phi \\
	\sin^2 \theta \cos \phi \sin \phi & \sin^2 \theta \sin^2 \phi & \sin \theta \cos \theta \sin \phi\\
	\sin \theta \cos \theta \cos \phi & \sin \theta \cos \theta \sin \phi & \cos^2\theta
\end{array}
\right).
\q

From Eq.~\eqref{PGW}, to obtain the radiated power of GWs, we need to compute the third derivative of $M_{ij}$. Notice that the components $M_{ij}$ depend on $\psi$, so one way to compute their derivatives is using $\dot{\phi}$ that is given by
\ee
\dot{\phi}=\frac{\tilde{L}}{ \mu R^2 \sin \theta}=\frac{\sqrt{\text{-C}} \csc\theta (e \cos (\phi  \sin\theta)-1)^2}{a^{3/2} \left(e^2-1\right)^{3/2} \sqrt{\mu }}.
\q
Then, using the chain rule, $\dot{M}_{ij}$ in terms of $\dot{\phi}$ is given by $\dot{M}_{ij}=\frac{dM_{ij}}{d\phi}\dot{\phi}$. Moreover, we can extend the same idea to the second and third derivatives. Therefore, we could arrive at
\ee
\dddot{M}_{ij}=\frac{\left(-C\right)^{3/2} \beta_{ij}}{a^{5/2} \left(e^2-1\right)^{5/2} \sqrt{\mu }},
\q
where $\beta_{ij}$ is a function of $e$, $\theta$, and $\phi$. The components of $\beta_{ij}$ are given by
\ee
\begin{aligned}
	\beta_{11}= &2\csc\theta \cos\phi (e \cos (\phi  \sin\theta)-1)^2 \\ &\times\left[4 \sin\phi (e \cos (\phi  \sin\theta)-1) \left(e \cos^2\theta \cos (\phi  \sin\theta)-1\right)+e \sin^3\theta \cos\phi \sin (\phi  \sin\theta)\right],\\
	\beta_{12}=&\beta_{21}=2\csc\theta (e \cos (\phi  \sin\theta)-1)^2  \Bigl[2 \sin^2\phi (e \cos (\phi  \sin\theta)-1) \left(e \cos^2\theta \cos (\phi  \sin\theta)-1\right) \\ &-2 \cos^2\phi (e \cos (\phi  \sin\theta)-1) \left(e \cos^2\theta \cos (\phi  \sin\theta)-1\right)+e \sin^3\theta \sin\phi \cos\phi \sin (\phi  \sin\theta)\Bigr],\\
	\beta_{13}=&\beta_{31}=\cot\theta \csc\theta (e \cos (\phi  \sin\theta)-1)^2 \\ &\times\left[2 e \sin^3\theta \cos\phi \sin (\phi  \sin\theta)+\sin\phi (e \cos (\phi  \sin\theta)-1) (e (2 \cos (2 \theta )-1) \cos (\phi  \sin\theta)-1)\right], \\
	\beta_{22}=&2\csc\theta \sin\phi (e \cos (\phi  \sin\theta)-1)^2 \\ &\times \left[e \sin^3\theta \sin\phi \sin (\phi  \sin\theta)-4 \cos\phi (e \cos (\phi  \sin\theta)-1) \left(e \cos^2\theta \cos (\phi  \sin\theta)-1\right)\right],\\
	\beta_{23}=&\beta_{32}=\cot\theta \csc\theta (e \cos (\phi  \sin\theta)-1)^2 \\ &\times \left[2 e \sin^3\theta \sin\phi \sin (\phi  \sin\theta)+\cos\phi (e \cos (\phi  \sin\theta)-1) ((e-2 e \cos (2 \theta )) \cos (\phi  \sin\theta)+1)\right], \\
	\beta_{33}=&2 e \cos^2\theta \sin (\phi  \sin\theta) (e \cos (\phi  \sin\theta)-1)^2.
\end{aligned}
\q
Using Eq.~\eqref{PGW}, we can obtain the radiated power of GWs, $\frac{dE_{\mathrm{GW}}^{\mathrm{quad}}}{dt}$, and the total energy loss due to gravitational quadrupole radiation, $\Delta E_{\mathrm{GW}}$,
\ee
\frac{dE_{\mathrm{GW}}^{\mathrm{quad}}}{dt}=\frac{C^3}{a^5 \left(e^2-1\right)^5 \mu } \mathcal{G}_1 (\theta, \phi, e),
\q
\ee
\Delta E_{\mathrm{GW}}^{\mathrm{quad}}=\int^{\infty}_{-\infty} \frac{dE_{\mathrm{GW}}^{\mathrm{quad}}}{dt} dt=\int^{2\pi/\sin\theta-\phi_1}_{\phi_1} \frac{dE_{\mathrm{GW}}^{\mathrm{quad}}}{dt} (\frac{d \phi}{dt})^{-1} d\phi=-\frac{(-C)^{5/2}}{a^{7/2} \left(e^2-1\right)^{7/2} \sqrt{\mu }} \mathcal{G}_2 (\theta,e),
\q
where 
\ee
\a
\mathcal{G}_1&=\frac{\csc^4\theta}{120} (e \cos (\phi  \sin\theta)-1)^4 \Bigl\{9 e^4 \cos (4 \phi  \sin\theta)+12 e \Bigl[e \left(\left(3 e^2+35\right) \cos (2 \phi  \sin\theta)-7 e \cos (3 \phi  \sin\theta)\right) \\
&-21 \left(e^2+4\right) \cos (\phi  \sin\theta)\Bigr]-2 e \cos (4 \theta ) \Bigl[e (3 e^2 \cos (4 \phi  \sin\theta)+2 \left(6 e^2+43\right) \cos (2 \phi  \sin\theta) \\
&+9 e^2-30 e \cos (3 \phi  \sin\theta)+82)-18 \left(5 e^2+4\right) \cos (\phi  \sin\theta)\Bigr]+\cos (2 \theta ) \Bigl[3 e^4 \cos (4 \phi  \sin\theta) \\
&+4 e \left(6 \left(28-3 e^2\right) \cos (\phi  \sin\theta)+ \left(3 e^3-26e\right) \cos (2 \phi  \sin\theta)-6 e^2 \cos (3 \phi  \sin\theta)\right) \\
&+9 e^4-136 e^2-360\Bigr]+27 e^4+444 e^2+408\Bigr\},
\b
\q
\ee
\a
\mathcal{G}_2&=\frac{\csc^4\theta}{1800} \Bigl\{-15 \arccos(e^{-1}) \Bigl[2 \left(15 e^4+323 e^2+308\right) e^2 \cos (4 \theta )+\left(-15 e^6+26 e^4+1976 e^2+720\right) \\
&\times \cos (2 \theta )-3 \left(15 e^6+404 e^4+1104 e^2+272\right)\Bigr]+450 \pi  e^6 \cos (4 \theta )+4926 \sqrt{e^2-1} e^4 \cos (4 \theta ) \\
&+9690 \pi  e^4 \cos (4 \theta )+13658 \sqrt{e^2-1} e^2 \cos (4 \theta )+9240 \pi  e^2 \cos (4 \theta )+796 \sqrt{e^2-1} \cos (4 \theta ) \\
&+(-225 \pi  e^6-2079 \sqrt{e^2-1} e^4+390 \pi  e^4+20368 \sqrt{e^2-1} e^2+29640 \pi  e^2+22316 \sqrt{e^2-1} \\
&+10800 \pi ) \cos (2 \theta )-675 \pi  e^6-7005 \sqrt{e^2-1} e^4-18180 \pi  e^4-47130 \sqrt{e^2-1} e^2 \\
&-49680 \pi  e^2-26640 \sqrt{e^2-1} -12240 \pi \Bigr\}.
\b
\q
Now, let us calculate the emission of electromagnetic dipole and quadrupole radiation due to the electric and magnetic charges on the orbit \eqref{R}. According to Ref.~\cite{Liu:2020vsy} and Appendix \ref{appendix}, the energy emission rate due to electromagnetic dipole radiation $P_{\mathrm{EM}}$ is
\m \label{P-EM}
\frac{dE_{\mathrm{EM}}^{\mathrm{dip}}}{dt}=-\frac{2 \mu^{2}((\Delta \sigma_q)^{2}+(\Delta \sigma_g)^{2})}{3} \ddot{R}^i \ddot{R}_i,
\n
where $\Delta \sigma_q=q_{2} / m_{2}-q_{1} / m_{1}$ and $\Delta \sigma_g=g_{2} / m_{2}-g_{1} / m_{1}$ are the dipole moments of electric charges and magnetic charges. Hence, the radiated power of electromagnetic waves, $\frac{dE_{\mathrm{EM}}^{\mathrm{dip}}}{dt}$, and the total energy loss due to electromagnetic radiation, $\Delta E_{\mathrm{GW}}$, are given by
\ee
\a
\frac{dE_{\mathrm{EM}}^{\mathrm{dip}}}{dt}=-\frac{\del C^2}{ a^4 \left(e^2-1\right)^4} \mathcal{K}_1 (\theta, \phi, e)
\b
\q
\ee
\Delta E_{\mathrm{EM}}^{\mathrm{dip}}=\int^{\infty}_{-\infty} \frac{dE_{\mathrm{EM}}^{\mathrm{dip}}}{dt} dt=\int^{2\pi/\sin\theta-\phi_1}_{\phi_1} \frac{dE_{\mathrm{EM}}^{\mathrm{dip}}}{dt} (\frac{d \phi}{dt})^{-1} d\phi= -\frac{\del (-C)^{3/2} \sqrt{\mu }}{ a^{5/2} \left(e^2-1\right)^{5/2}} \mathcal{K}_2 (\theta,e),
\q
where
\ee
\a
\mathcal{K}_1&=\frac{\csc^2\theta}{12} (e \cos (\phi  \sin\theta)-1)^4 \Bigl(2 e^2 \cos (2 \phi  \sin\theta)+e^2 \cos (2\phi  \sin\theta+2\theta ))+e^2 \cos (2 \theta -2 \phi  \sin\theta) \\
&+2 e^2 \cos (2 \theta )+2 e^2-8 e \cos (\phi  \sin\theta)-4 e \cos (\phi  \sin\theta+2 \theta )-4 e \cos (2 \theta -\phi  \sin\theta)+8\Bigr),
\b
\q
\ee
\a
\mathcal{K}_2&=\frac{\csc^2\theta}{36} \Bigl(-3 \arccos(e^{-1}) \left(\left(3 e^2+20\right) e^2 \cos (2 \theta )+3 e^4+28 e^2+16\right) +(9 \pi  e^4+55 \sqrt{e^2-1} e^2 \\
&+60 \pi  e^2+14 \sqrt{e^2-1}) \cos (2 \theta )+9 \pi  e^4 +55 \sqrt{e^2-1} e^2+84 \pi  e^2+86 \sqrt{e^2-1} +48 \pi \Bigr).
\b
\q
From Appendix \ref{appendix}, the relation between electromagnetic quadrupole radiation and gravitational quadrupole radiation is given by 
\ee
\frac{dE_{\mathrm{EM}}^{\mathrm{quad}}}{dt}\equiv\frac{\mu^{2}((q_2/m_2^2+q_1/m_1^2)^{2}+(g_2/m_2^2+g_1/m_1^2)^{2})}{4} \frac{dE_{\mathrm{GW}}^{\mathrm{quad}}}{dt}.
\q
Furthermore, the contribution of the quadrupole term of electromagnetic radiation is always smaller than the quadrupole term of gravitational radiation. Therefore, the total energy loss due to electromagnetic dipole and quadrupole radiation and gravitational quadrupole radiation is given by
\ee
\Delta E=\Delta E_{\mathrm{EM}}^{\mathrm{dip}}+\Delta E_{\mathrm{EM}}^{\mathrm{quad}}+\Delta E_{\mathrm{GW}}^{\mathrm{quad}}=\Delta E_{\mathrm{EM}}^{\mathrm{dip}}+\left(1+ \Lambda \right) \Delta E_{\mathrm{GW}}^{\mathrm{quad}},
\q
where $\Lambda=\frac{\mu^{2}((q_2/m_2^2+q_1/m_1^2)^{2}+(g_2/m_2^2+g_1/m_1^2)^{2})}{4}$. In this Section, we have calculated the total emission rates of energy due to gravitational and electromagnetic radiation from BBHs with electric and magnetic charges in hyperbolic orbits. In the universe, the two-body dynamical capture is an absolutely common and effective way to form BBH systems. We will derive the merger rate of BBHs with electric and magnetic charges from the two-body dynamical capture in the next Section.

\section{\label{III}Merger rate of BBHs with electric and magnetic charges from the two-body dynamical capture}
If two BHs with electric and magnetic charges are getting closer and closer, the total energy loss due to gravitational and electromagnetic radiation could exceed the orbital kinetic energy. Hence the unbound system cannot escape to infinity anymore and will form a bound binary with negative orbital energy. Therefore, this binary immediately merges through consequent large electromagnetic and gravitational radiation. For such a process, we can estimate the cross section and calculate the merger rate of BBHs with electric and magnetic charges from the two-body dynamical capture.

Let us consider the interaction of two dyonic BHs with masses and charges ($m_1$, $q_1$, $g_1$) and ($m_2$, $q_2$, $g_2$), and assume that they have an initial relative velocity $v$, the impact parameter $b$ and the distance of periastron $r_p$. According to the definition of $r_p$ and the orbit \eqref{R}, we have $r_p\equiv R_{min}=R(\phi=\frac{\pi}{\sin \theta})=a(e-1)$. We could approximate the trajectory of a close encounter by the hyperbolic with $e\rightarrow 1$ since when the two dyonic BHs pass by closer and closer, the true trajectory is physically indistinguishable from a parabolic one near the periastron where electromagnetic and gravitational radiation dominantly occurs. According to Sec. \ref{II}, the total energy loss due to electromagnetic radiation and gravitational  radiation by the close encounter can be evaluated by using $e \rightarrow 1$ and the periastron $r_p \equiv a(e-1)$, namely
\ee
\a
\Delta E_{\mathrm{EM}}^{\mathrm{dip}}+\left(1+ \Lambda \right) \Delta E_{\mathrm{GW}}^{\mathrm{quad}}
\b
\q
where
\ee
\Delta E_{\mathrm{EM}}^{\mathrm{dip}}= -\frac{\pi \del (-C)^{3/2} \sqrt{\mu } (23 \cos (2 \theta )+47) \csc ^2(\theta )}{48 \sqrt{2} r_p^{5/2}},
\q
\ee
\Delta E_\mathrm{{GW}}^{\mathrm{quad}}=-\frac{\pi  (-C)^{5/2} (-2707 \cos (2 \theta )-1292 \cos (4 \theta )+5385) \csc^4\theta}{960 \sqrt{2} \sqrt{\mu } r_p^{7/2}}.
\q
The definition of the impact parameter $b$ is the distance from the origin to the asymptotes of the hyperbolic orbit. One asymptote of the hyperbolic orbit is given by
\ee
\a
\label{xyz}
x&=\lim _{\phi \rightarrow \phi_1} \left( \frac{a\left(e^{2}-1\right)}{1-e \cos (\phi \sin \theta)} \sin \theta \cos \phi \right)-l\sin\theta\cos\phi_1, \\
y&=\lim _{\phi \rightarrow \phi_1} \left( \frac{a\left(e^{2}-1\right)}{1-e \cos (\phi \sin \theta)} \sin \theta \sin \phi \right)-l \sin \theta \sin \phi_1, \\
z&=\lim _{\phi \rightarrow \phi_1} \left( \frac{a\left(e^{2}-1\right)}{1-e \cos (\phi \sin \theta)} \cos \theta \right)-l \cos \theta.
\b
\q
Here, when choosing $y_{0}=0$, we can get the solution of $l_0$,
\ee
\a
\label{l}
l=\lim _{\phi \rightarrow \phi_1} \left(\frac{a\left(e^{2}-1\right)}{1-e \cos (\phi \sin \theta)} \frac{\sin \phi}{\sin \phi_1} \right).
\b
\q
From Eqs.~\eqref{xyz} and \eqref{l}, when $y_{0}=0$, $x_0$ and $z_0$ are expressed as
\ee
\a
x_0&=\lim _{\phi \rightarrow \phi_1} \left( \frac{a\left(e^{2}-1\right)}{1-e \cos (\phi \sin \theta)} \sin \theta \cos \phi-\frac{a\left(e^{2}-1\right)}{1-e \cos (\phi \sin \theta)} \frac{\sin \phi}{\sin \phi_1 }\sin\theta\cos\phi_1 \right) \\
&=-a \sqrt{e^2-1} \csc \left(\arccos(e^{-1}) \csc\theta\right),\\
z_0&=\lim _{\phi \rightarrow \phi_1} \left( \frac{a\left(e^{2}-1\right)}{1-e \cos (\phi \sin \theta)} \cos \theta -\frac{a\left(e^{2}-1\right)}{1-e \cos (\phi \sin \theta)} \frac{\sin \phi}{\sin \phi_1 }\cos \theta \right) \\
&=-a \sqrt{e^2-1} \cot\theta \cot \left(\arccos(e^{-1}) \csc\theta\right).
\b
\q
Therefore, we get the point of intersection of the asymptotes and $x$-$z$ plane, $(x_0,0,z_0)$. We denote the vector $\vec{v_1}$ as $\vec{v}_1=(x_0,0,z_0)$. Using the unit vector along the asymptotes of the hyperbolic orbit, $\vec{v}_2=\left(\sin \theta  \cos \phi_1,\sin \theta  \sin \phi_1,\cos \theta \right)$, the impact parameter $b$ is given by
\ee
\a
\label{b}
b=\sqrt{\vec{v}_1\cdot \vec{v}_1-(\vec{v}_1\cdot\vec{v}_2)^2}=a \sqrt{e^2-1},
\b
\q
which is independent of $\theta$. It shows that no matter the orbit is three-dimensional ($\theta\neq\pi/2$) or two-dimensional ($\theta=\pi/2$), we always have $b=a \sqrt{e^2-1}$.
From Eq.~\eqref{b} and $r_p \equiv a(e-1)$, we can get $b^2=r_p^2+2 a r_p$. Note that the total energy can be expressed as $E \equiv -\frac{C}{2a}=\frac{\mu v^2}{2} $, so $a=-\frac{C}{\mu v^2}$. Therefore, we find the relation between $r_p$ and $b$ as
\ee
b^2=r_p^2 - \frac{2C r_p}{\mu v^2}.
\q
In the limit of a strong electromagnetic and gravitational radiation focusing, i.e. $r_p\ll b$, then we get the distance of the closest approach $r_p$ as
\ee
\label{rp}
r_p=- \frac{\mu v^2 b^2}{2C}.
\q
The condition for the dyonic BHs to form a bound system is that the total energy loss due to electromagnetic and gravitational radiation is larger than the kinetic energy $\mu v^2/2$, i.e.,
\ee
\label{Condition}
\Delta E+\frac{\mu v^2}{2}<0.
\q
From Eqs.~\eqref{rp} and \eqref{Condition}, we could obtain the merging cross-section $\sigma=\pi b_{\max}^2$, where $b_{\max}$ is the maximum impact parameter for the dyonic BHs to form a bound system and is determined by
\ee
\a
\label{bb}
&\frac{\pi  (\Lambda +1) C^6 (5385-2707 \cos (2 \theta )-1292 \cos (4 \theta )) \csc ^4\theta}{120 b_{\max}^7 \mu ^4 v^7} \\
+&\frac{\pi  \del  C^4 (23 \cos (2 \theta )+47) \csc ^2\theta}{12 b_{\max}^5 \mu ^2 v^5}=\frac{\mu v^2}{2}.
\b
\q

\iffalse

When  the total quadrupole radiation is dominant (i.e. $ (1+\La) \Delta E_{\mathrm{GW}}^{\mathrm{quad}}>\Delta E_{\mathrm{EM}}^{\mathrm{dip}} $ ), the cross-section of merging is given by
\ee
\a
\label{quad}
\sigma_{quad}\approx \frac{\pi ^{9/7} (\Lambda +1)^{2/7} (-C)^{12/7} (-2707 \cos (2 \theta )-1292 \cos (4 \theta )+5385)^{2/7} \csc ^{\frac{8}{7}}(\theta )}{2^{4/7} 15^{2/7} \mu ^{10/7} v^{18/7}},
\b
\q
which is consistent with \cite{Mouri:2002mc}, and when the dipole radiation is dominant (i.e. $(1+\La) \Delta E_{\mathrm{GW}}^{\mathrm{quad}}<\Delta E_{\mathrm{EM}}^{\mathrm{dip}} $ ), the cross-section of merging is given by
\ee
\label{dipole}
\sigma_{dip}\approx \frac{\pi ^{7/5} \del ^{2/5} (-C)^{8/5} (23 \cos (2 \theta )+47)^{2/5} \csc ^{\frac{4}{5}}(\theta )}{6^{2/5} \mu ^{6/5} v^{14/5}}.
\q
Noting that the Schwarzschild radius $\sigma_{Sch} \simeq \pi (2G m)^2 \simeq (v)^{18/7} \sigma_{quad}$, $v \ll 1$ and $\sigma \geq \sigma_{quad}$; therefore, no matter whether gravitational radiation is dominant or electromagnetic radiation is dominant, the Newtonian approximation is sufficiently accurate.
\fi

Here, we want to reminder the reader that $\theta$ is a function of $b_{\max}$ and is given by $\cos^2\theta=\frac{D^2}{D^2+\mu^2 v^2 b_{\max}^2}$.
Finally, we could achieve the differential merger rate of dyonic BHs from the two-body dynamical capture,
\ee
\label{merger rate}
dR=n(m_1,q_1,g_1)n(m_2,q_2,g_2)\left\langle\sigma v\right\rangle dm_1dm_2dq_1dq_2dg_1dg_2
\q
where $\left\langle\sigma v\right\rangle$ denotes the average over relative velocity distribution with $\sigma=\pi b_{\max}^2$ in Eq. \eqref{b} and $n(m_1,q_1,g_1)$ and $n(m_2,q_2,g_2)$ are the comoving average number density of dyonic BHs with mass, electric and magnetic charges ($m_1$, $q_1$, $g_1$) and ($m_2$, $q_2$, $g_2$). In this section, we develop a formalism to derive the merger rate of BBHs with electric and magnetic charges from the two-body dynamical capture. Next, we will apply the formalism to find the effects of the charges on the merger rate for the near-extremal case.

\section{\label{IV}Effects of the charges on the merger rate for the near-extremal case}

The origin of those BHs with electric and magnetic charges may be the primordial BH (PBH) which are BHs formed in the radiation-dominated era of the early Universe due to the collapse of large energy density fluctuations \cite{Zeldovich:1967lct, Hawking:1971ei, Carr:1974nx}. A PBH can be magnetized by the accretion of monopoles in the early Universe. For instance, PBHs in a strong magnetic field can produce a pair of magnetic monopoles through pair production \cite{Das:2021wei,Kobayashi:2021des}. The magnetized PBH can also produce a strong magnetic field, which results in the accretion of electric charges \cite{Wald:1974np}. This is the mechanism for BHs to have electric and magnetic charges in the early Universe. Recently, Ref.~\cite{Kritos:2021nsf} shows that PBHs can be near-extremal charged. In this section, we will apply the formalism developed in the last section to find the effects of the charges on the merger rate for the near-extremal case. For simplicity, we consider a special model assuming that all PBHs have the same mass $M_{\PBH}$. In such a model, the number density of dyonic PBHs is given by
\ee
n(m,q,g)=\frac{1}{4}\frac{\fpbh\rho_{\mathrm{DM}}}{M_{\PBH}} \left(\delta(\frac{q}{m}-\iota_e)+\delta(\frac{q}{m}+\iota_e)\right)\left(\delta(\frac{g}{m}-\iota_m)+\delta(\frac{g}{m}+\iota_m)\right) \delta(m-M_{\PBH}),
\q
where $\rho_{\mathrm{DM}}$ is the dark matter energy density at present, and $\fpbh$ is the fraction of PBHs in the dark matter. This model suggests that the part of the Universe under study is charge neutral, but charges are locally separated by some mechanisms found in Refs.~\cite{Wald:1974np, Bai:2019zcd}.
Here, we only consider the near-extremal case. In other words, we take $\iota_m\equiv\sqrt{1-\iota_e ^2}$. For simplicity, we choose $\iota_e\geq0$. In the calculation,  we take the Maxwell-Boltzmann distribution $P(v) \propto v^{2} \exp \left(-v^{2} / v_{0}^{2}\right)$ for the velocity distribution of BHs with the most probable velocity $v_{0}=100$ km/s. For the charge-neutral Schwarzschild BHs, the merger rate of PBH binaries from the two-body capture is $R_{\mathrm{Sch}}\approx 1.5\times 10^{-8} \fpbh^{2} \mathrm{Gpc}^{-3} \mathrm{yr}^{-1}$ that is independent of $M_{\PBH}$ and scales as $\fpbh^2$.

\begin{figure}[htbp!]
\includegraphics[width=0.45\textwidth]{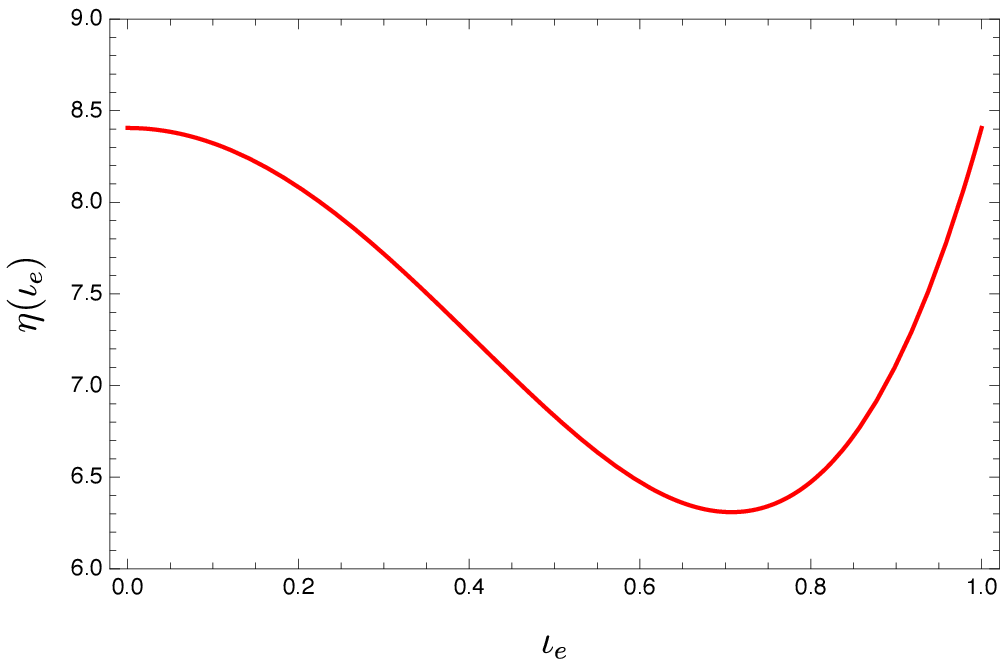}
\includegraphics[width=0.485\textwidth]{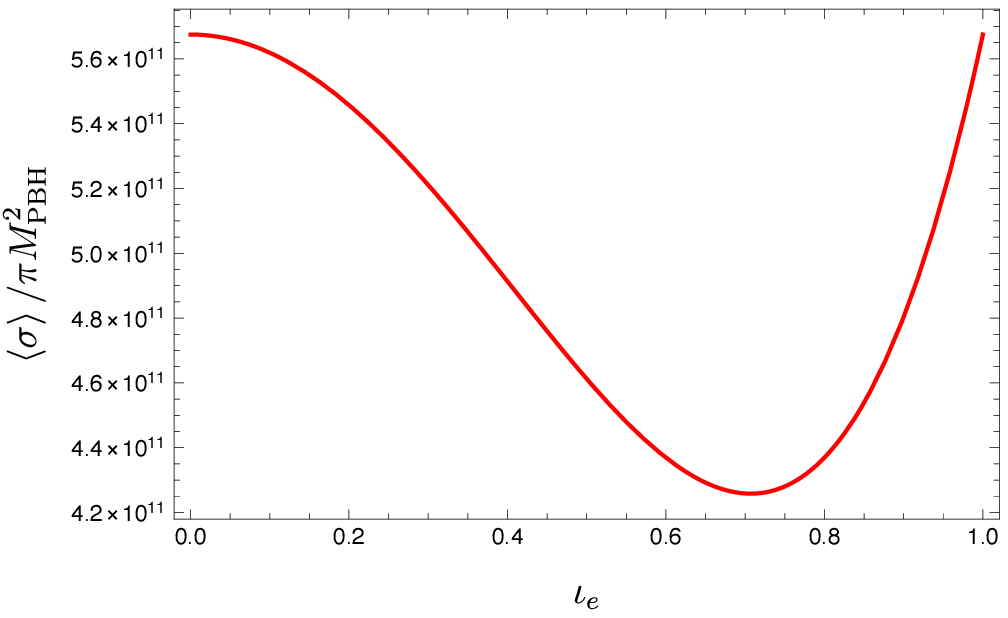}
\caption{\label{fig:2} Left: The plot of $\eta(\iota_e)$ as a function of $\iota_e$. Right: The plot of $\left\langle \sigma \right\rangle/\pi M_ {\mathrm{PBH}}^2$ as a function of $\iota_e$.}
\end{figure}

To show the effects of charges $\iota_e$ on the merger rate of PBH binaries from the two-body dynamical capture for the near-extremal case, we define a function of $\iota_e$ as
\ee
\label{eta}
\eta(\iota_e)\equiv\frac{R(\iota_e)}{R_{\mathrm{Sch}}},
\q
where $R(\iota_e)$ is the total merger rate of near-extremal PBH binaries with electric charge-to-mass ratio $\iota_e$ and $R_{\mathrm{Sch}}$ is the total merger rate of PBH binaries in charge-neutral case. In this special model, $C$ equals to $-M_{\PBH}^2\left(1\pm \iota_e^2 \pm \left(1-\iota_e^2\right)\right)$, and $D$ equals to $0,\pm 2 \iota_e  \sqrt{1-\iota_e ^2} M_{\PBH}^2$ in different cases as shown in Table \ref{table1}. The total merger rate of near-extremal PBH binaries, $R(\iota_e)$, is the sum of the merger rate of different cases. Notice that from Eq.~\eqref{bb} it follows $b_{\max}\propto M_{\PBH}$ and $\sigma \propto M_{\PBH}^2$, we find that $R(\iota_e)$  is independent of $M_{\PBH}$ and scales as $\fpbh^2$. Therefore, $\eta(\iota_e)$ is independent of $M_{\PBH}$ and $\fpbh$ and is only a function of $\iota_e$.  In Fig.~\ref{fig:2}, we plot $\eta(\iota_e)$ as the function of $\iota_e$. From the definition, we find $\eta(\iota_e)=\eta(\sqrt{1-\iota_e^2})$ and show that $\eta(\iota_e)$ decreases as $\iota_e$ increases and reaches the minimum value of $\eta(\frac{\sqrt{2}}{2}) \approx 6.3 $ in $\iota_e \in [0,\frac{\sqrt{2}}{2}]$. For $\iota_e \in [\frac{\sqrt{2}}{2},1]$, $\eta(\iota_e)$ increases as $\iota_e$ increases and reaches the maximum value of $\eta(1) \approx 8.4$. As shown in Fig.~\ref{fig:2}, the effects of the charges on the merger rate for the near-extremal case cannot be ignored. {In Fig.~\ref{fig:2}, we also show that the averaged merging cross section is always much larger than that corresponding to the event horizon radius. Therefore, the Newtonian approximation is sufficiently accurate.}

%$C$ could equal $-M_{\PBH}^2\left(1\pm \iota_e^2 \pm \left(1-\iota_e^2\right)\right)$  represent the PBH binary has the opposite (same) electric charges and opposite (same) magnetic charges, respectively.

\begin{center}
\begin{table}[!h]
        \centering

\begin{tabular}{|l|l|l|l|l|}
\hline
  &$q_1=q_2 \geq0$  & $q_1=-q_2 \geq 0$ & $q_1=q_2 <0$  & $q_1=-q_2 <0$   \\
\hline
 \multirow{2}{*}{$g_1=g_2 \geq0$} & $C=0$  & $C=-2 \iota_e ^2 M_{\PBH}^2$ & $C=0$  & $C=2 \iota_e ^2 M_{\PBH}^2$\\
 & $D=0$  & $D=-2 \iota_e  \iota_m M_{\PBH}^2$ & $D=0$  & $D=-2 \iota_e  \iota_m M_{\PBH}^2$\\
\hline
 \multirow{2}{*}{$g_1=-g_2 \geq0$} & $C=-2 \iota_m^2 M_{\PBH}^2$  & $C=-2  M_{\PBH}^2$ & $C=-2 \iota_m^2 M_{\PBH}^2$  & $C=-2  M_{\PBH}^2$ \\
 & $D=2 \iota_e  \iota_m M_{\PBH}^2$  & $D=0$ & $D=-2 \iota_e  \iota_m M_{\PBH}^2$  & $D=0$ \\
\hline
 \multirow{2}{*}{$g_1=g_2<0$} & $C=0$   & $C=-2 \iota_e ^2 M_{\PBH}^2$ & $C=0$ & $C=-2 \iota_e ^2 M_{\PBH}^2$ \\
 & $D=0$   & $D=2 \iota_e  \iota_m M_{\PBH}^2$ & $D=0$ & $D=-2 \iota_e  \iota_m M_{\PBH}^2$ \\
\hline
  \multirow{2}{*}{$g_1=-g_2<0$} & $C=-2 \iota_m^2 M_{\PBH}^2$ & $C=-2  M_{\PBH}^2$  & $C=-2 \iota_m^2 M_{\PBH}^2$  &  $C=-2  M_{\PBH}^2$  \\
  & $D=-2 \iota_e  \iota_m M_{\PBH}^2$ & $D=0$  & $D=2 \iota_e  \iota_m M_{\PBH}^2$  &  $D=0$ \\
 \hline
\end{tabular}
\caption{\label{table1}The value of $C$ and $D$ in different cases.}
%Here, we have defined $\iota_m\equiv\sqrt{1-\iota ^2}$.
        \end{table}
\end{center}

\section{\label{Con}Conclusion}

In this work, we have derived the hyperbolic orbit of BBHs with electric and magnetic charges. In the low-velocity and weak-field regime, by using a Newtonian method, we calculate the total emission rate of energy due to gravitational and electromagnetic radiation from BBHs with electric and magnetic charges in hyperbolic orbits. We also develop a formalism to derive the merger rate of BBHs with electric and magnetic charges from the two-body dynamical capture. We apply this formalism to estimate the effects of the charges on the merger rate for the near-extremal case and find that the effects cannot be ignored. 
%The results of this work can be used to investigate whether BHs have electric and magnetic charges or not.

{In our calculation, we don't assume the mass of binary. For the solar mass range, combined with Ref.~\cite{Liu:2020vsy, Liu:2020bag}, the results of this work could provide rich information and a crosscheck to test whether LIGO-Virgo-KAGRA black holes have electric and magnetic charges. On the other hand, those BBHs whose mass is smaller than one solar masses must be PBHs instead of astrophysical BHs. Those extremal charged PBHs are stable and could account for all dark matter without requiring physics beyond the standard model, even though there are many constraints for uncharged PBHs as dark matter \cite{Sasaki:2018dmp, Carr:2020gox, Carr:2020xqk}. Two extremal-charged PBHs with opposite charges could form a bound system through the two-body capture. When they merge, the burst of gamma rays due to the annihilation of charges could be detected by observations.}

We find that the BHs with electric and magnetic charges can form a bound system due to gravitational and electromagnetic radiation. Another possibility is that charged BHs do not end up with a bound system in a single encounter but inspiral and enter another scattering event through a hyperbolic encounter. For two BHs with electric and magnetic charges, if the relative velocity or distance is large enough, then the two-body capture cannot happen. These events, however, can generate bursts of GWs. Compared with the GW burst produced by the encounter of Schwarzschild BHs,  the GW burst produced by the encounter of BHs with electric and magnetic charges has different characteristics and phenomena. The characteristic peak frequency and detection of such GW bursts is an interesting issue, and we will leave this topic for future work.

\acknowledgments
%We thank the referee for providing constructive comments and suggestions to improve the quality of this paper.
S.P.K. is supported by the National Research Foundation of Korea (NRF) funded by the Ministry of Education (2019R1I1A3A01063183).
L.L. is supported by the National Natural Science Foundation of China (Grant No.~12247112 and No.~12247176).
Z.-C.C. is supported by the National Natural Science Foundation of China (Grant No.~12247176 and No.~12247112) and the China Postdoctoral Science Foundation Fellowship No.~2022M710429.

\appendix
\section{\label{appendix}Electromagnetic dipole and quadrupole radiation from dyonic BBHs}

When the differences $\Delta \sigma_q$ of electric charge-to-mass ratios and $\Delta \sigma_g$ of magnetic charge-to-mass ratios are very small or even vanishes, the charge quadrupole might be extremely important. Here, we will consider the electromagnetic dipole and quadrupole radiation from dyonic BBHs. We first derive the emission of electromagnetic radiation from electric charges, then calculate the emission from magnetic charges, and finally superimpose their fields.

Following Ref.~\cite{Liu:2022wtq}, the energy  emission due to electromagnetic dipole and quadrupole radiation is given by
\ee
\frac{dE_{\mathrm{EM}}}{dt}|_{e}=\frac{dE_{\mathrm{EM}}^{\mathrm{dip}}}{dt}|_{e}+\frac{dE_{\mathrm{EM}}^{\mathrm{quad}}}{dt}|_{e},
\q
and
\ee
\frac{dE_{\mathrm{EM}}^{\mathrm{dip}}}{dt} |_{e}=-\frac{2\ddot{p}^2}{3},
\q
\ee
\frac{dE_{\mathrm{EM}}^{\mathrm{quad}}}{dt} |_{e}=-\frac{\ddddot{D}_{ij}\ddddot{D}_{ij}}{20}.
\q
where $p^{i}=\mu \Delta \sigma_q R^i$ is electric charge dipole and $D^{i j}=\mu\left(\frac{Q_1}{m_1^2} +\frac{Q_2}{m_2^2}\right) Q^{i j}$ is traceless electric charge quadrupole.

An important consequence of the enhanced symmetry due to the existence of magnetic monopoles is that the classical dynamics of the charges, fields and Maxwell's equations are all invariant under the dual transformation,
\m
\begin{aligned}
&\boldsymbol{E}^{\prime}=\boldsymbol{E} \cos \alpha-\boldsymbol{B} \sin \alpha, \\
&\boldsymbol{B}^{\prime}=\boldsymbol{E} \sin \alpha+\boldsymbol{B} \cos \alpha,\\
&q^{\prime} = q \cos \alpha+g \sin \alpha,\\
&g^{\prime} = g \cos \alpha-g \sin \alpha.
\end{aligned}
\n
Choosing $\alpha=\pi / 2$, pure electric charges could transform to pure magnetic charges. This helps us to directly find the fields emanating from magnetic charges from the results for pure electric charges. For $\alpha=\pi / 2$, it is easy to find $\boldsymbol{E}_{2} \propto \boldsymbol{B}_{1}$ and $\boldsymbol{B}_{2} \propto  \boldsymbol{E}_{1}$ by labeling the fields from the electric charge, $\boldsymbol{E}_{1}, \boldsymbol{B}_{1},$ and those from the dual transformation, $\boldsymbol{E}_{2}, \boldsymbol{B}_{2}$. Now, we consider the integrated energy density on a shell for electric and magnetic fields. Notice that the electric dipole and quadruple have the same direction as the magnetic dipole and quadruple. Therefore, we get $\boldsymbol{E}_{1} \perp \boldsymbol{E}_{2}$, $\boldsymbol{B}_{1} \perp \boldsymbol{B}_{2}$, $\boldsymbol{E}_{1} ||\boldsymbol{B}_{2}$, and $\boldsymbol{E}_{2} ||\boldsymbol{B}_{1}$. We then consider the result for the energy density and momentum density
\ee
u=\frac{1}{2}\left(E^{2}+B^{2}\right)
=\frac{1}{2}(E_{1}^{2}+B_{1}^{2}+ E_{2}^{2}+B_{2}^{2}+2\left(\boldsymbol{E}_{1} \cdot \boldsymbol{E}_{2}+\boldsymbol{B}_{1} \cdot \boldsymbol{B}_{2}\right))=u_1+u_2,
\q
\ee
\boldsymbol{\mathcal{P}}=\boldsymbol{E} \times \boldsymbol{B}=\boldsymbol{E}_{1} \times \boldsymbol{B}_{1}+\boldsymbol{E}_{2} \times \boldsymbol{B}_{2}
+\boldsymbol{E}_{1} \times \boldsymbol{B}_{2}+\boldsymbol{E}_{2} \times \boldsymbol{B}_{1}=\boldsymbol{\mathcal{P}_1}+\boldsymbol{\mathcal{P}_2}.
\q
Using $\frac{dE_{\mathrm{EM}}}{dt}=- r^2 \int d\Omega \hat{r}\cdot \boldsymbol{\mathcal{P}}$, we get $\frac{dE_{\mathrm{EM}}}{dt}=\frac{dE_{\mathrm{EM}}}{dt}|_{e}+\frac{dE_{\mathrm{EM}}}{dt}|_{m}$. This means that the total energy emissions due to electromagnetic dipole and quadrupole radiation are given by
\ee
\frac{dE_{\mathrm{EM}}}{dt}=\frac{dE_{\mathrm{EM}}^{\mathrm{dip}}}{dt}+\frac{dE_{\mathrm{EM}}^{\mathrm{quad}}}{dt},
\q
where
\ee
\frac{dE_{\mathrm{EM}}^{\mathrm{dip}}}{dt}=\frac{dE_{\mathrm{EM}}^{\mathrm{dip}}}{dt}|_e+\frac{dE_{\mathrm{EM}}^{\mathrm{dip}}}{dt}|_m=-\frac{2 \mu^{2}((q_2/m_2-q_1/m_1)^{2}+(g_2/m_2-g_1/m_1)^{2})}{3} \ddot{R}^i \ddot{R}_i,
\q
\ee
\frac{dE_{\mathrm{EM}}^{\mathrm{quad}}}{dt}=\frac{dE_{\mathrm{EM}}^{\mathrm{quad}}}{dt}|_e+\frac{dE_{\mathrm{EM}}^{\mathrm{quad}}}{dt}|_m=-\frac{\mu^2 ((q_2/m_2^2+q_1/m_1^2)^{2}+(g_2/m_2^2+g_1/m_1^2)^{2})}{20}\dddot{Q}_{ij}\dddot{Q}_{ij}.
\q
Notice that $\frac{dE_{\mathrm{GW}}^{\mathrm{quad}}}{dt} \equiv  -\frac{1}{5}\left(\dddot{Q}_{ij}\dddot{Q}_{ij}\right)$, we obtain the relation between electromagnetic quadrupole radiation and gravitational quadrupole radiation,
\ee
\frac{dE_{\mathrm{EM}}^{\mathrm{quad}}}{dt}\equiv\frac{\mu^{2}((q_2/m_2^2+q_1/m_1^2)^{2}+(g_2/m_2^2+g_1/m_1^2)^{2})}{4} \frac{dE_{\mathrm{GW}}^{\mathrm{quad}}}{dt}.
\q
From the metric constrains, $q_1^2+g_1^2 \leq m_1^2$ and $q_2^2+g_2^2 \leq m_2^2$, it is straightforward to prove that  $\frac{dE_{\mathrm{EM}}^{i,\mathrm{quad}}}{dt}\leq \frac{1}{4} \frac{dE_{\mathrm{GW}}^{\mathrm{quad}}}{dt}$ is always hold.

%%%%%%%%%%%%%%%%%%%%%%%%%%%%%%%%%%%%%%%%
%%%%%%%%%%%%%%%%%%%%%%%%%%%%%%%%%%%%%%%%

\bibliography{Ref}

\begin{thebibliography}{58}%
\makeatletter
\providecommand \@ifxundefined [1]{%
 \@ifx{#1\undefined}
}%
\providecommand \@ifnum [1]{%
 \ifnum #1\expandafter \@firstoftwo
 \else \expandafter \@secondoftwo
 \fi
}%
\providecommand \@ifx [1]{%
 \ifx #1\expandafter \@firstoftwo
 \else \expandafter \@secondoftwo
 \fi
}%
\providecommand \natexlab [1]{#1}%
\providecommand \enquote  [1]{``#1''}%
\providecommand \bibnamefont  [1]{#1}%
\providecommand \bibfnamefont [1]{#1}%
\providecommand \citenamefont [1]{#1}%
\providecommand \href@noop [0]{\@secondoftwo}%
\providecommand \href [0]{\begingroup \@sanitize@url \@href}%
\providecommand \@href[1]{\@@startlink{#1}\@@href}%
\providecommand \@@href[1]{\endgroup#1\@@endlink}%
\providecommand \@sanitize@url [0]{\catcode `\\12\catcode `\$12\catcode
  `\&12\catcode `\#12\catcode `\^12\catcode `\_12\catcode `\%12\relax}%
\providecommand \@@startlink[1]{}%
\providecommand \@@endlink[0]{}%
\providecommand \url  [0]{\begingroup\@sanitize@url \@url }%
\providecommand \@url [1]{\endgroup\@href {#1}{\urlprefix }}%
\providecommand \urlprefix  [0]{URL }%
\providecommand \Eprint [0]{\href }%
\providecommand \doibase [0]{http://dx.doi.org/}%
\providecommand \selectlanguage [0]{\@gobble}%
\providecommand \bibinfo  [0]{\@secondoftwo}%
\providecommand \bibfield  [0]{\@secondoftwo}%
\providecommand \translation [1]{[#1]}%
\providecommand \BibitemOpen [0]{}%
\providecommand \bibitemStop [0]{}%
\providecommand \bibitemNoStop [0]{.\EOS\space}%
\providecommand \EOS [0]{\spacefactor3000\relax}%
\providecommand \BibitemShut  [1]{\csname bibitem#1\endcsname}%
\let\auto@bib@innerbib\@empty
%</preamble>
\bibitem [{\citenamefont {Abbott}\ \emph {et~al.}(2016)\citenamefont {Abbott}
  \emph {et~al.}}]{Abbott:2016blz}%
  \BibitemOpen
  \bibfield  {author} {\bibinfo {author} {\bibfnamefont {B.~P.}\ \bibnamefont
  {Abbott}} \emph {et~al.} (\bibinfo {collaboration} {LIGO Scientific,
  Virgo}),\ }\bibfield  {title} {\enquote {\bibinfo {title} {{Observation of
  Gravitational Waves from a Binary Black Hole Merger}},}\ }\href {\doibase
  10.1103/PhysRevLett.116.061102} {\bibfield  {journal} {\bibinfo  {journal}
  {Phys. Rev. Lett.}\ }\textbf {\bibinfo {volume} {116}},\ \bibinfo {pages}
  {061102} (\bibinfo {year} {2016})},\ \Eprint
  {http://arxiv.org/abs/1602.03837} {arXiv:1602.03837 [gr-qc]} \BibitemShut
  {NoStop}%
%%CITATION = ARXIV:1602.03837;%%
\bibitem [{\citenamefont {Abbott}\ \emph
  {et~al.}(2019{\natexlab{a}})\citenamefont {Abbott} \emph
  {et~al.}}]{LIGOScientific:2018mvr}%
  \BibitemOpen
  \bibfield  {author} {\bibinfo {author} {\bibfnamefont {B.~P.}\ \bibnamefont
  {Abbott}} \emph {et~al.} (\bibinfo {collaboration} {LIGO Scientific,
  Virgo}),\ }\bibfield  {title} {\enquote {\bibinfo {title} {{GWTC-1: A
  Gravitational-Wave Transient Catalog of Compact Binary Mergers Observed by
  LIGO and Virgo during the First and Second Observing Runs}},}\ }\href
  {\doibase 10.1103/PhysRevX.9.031040} {\bibfield  {journal} {\bibinfo
  {journal} {Phys. Rev.}\ }\textbf {\bibinfo {volume} {X9}},\ \bibinfo {pages}
  {031040} (\bibinfo {year} {2019}{\natexlab{a}})},\ \Eprint
  {http://arxiv.org/abs/1811.12907} {arXiv:1811.12907 [astro-ph.HE]}
  \BibitemShut {NoStop}%
%%CITATION = ARXIV:1811.12907;%%
\bibitem [{\citenamefont {Abbott}\ \emph
  {et~al.}(2021{\natexlab{a}})\citenamefont {Abbott} \emph
  {et~al.}}]{Abbott:2020niy}%
  \BibitemOpen
  \bibfield  {author} {\bibinfo {author} {\bibfnamefont {R.}~\bibnamefont
  {Abbott}} \emph {et~al.} (\bibinfo {collaboration} {LIGO Scientific,
  Virgo}),\ }\bibfield  {title} {\enquote {\bibinfo {title} {{GWTC-2: Compact
  Binary Coalescences Observed by LIGO and Virgo During the First Half of the
  Third Observing Run}},}\ }\href {\doibase 10.1103/PhysRevX.11.021053}
  {\bibfield  {journal} {\bibinfo  {journal} {Phys. Rev. X}\ }\textbf {\bibinfo
  {volume} {11}},\ \bibinfo {pages} {021053} (\bibinfo {year}
  {2021}{\natexlab{a}})},\ \Eprint {http://arxiv.org/abs/2010.14527}
  {arXiv:2010.14527 [gr-qc]} \BibitemShut {NoStop}%
\bibitem [{\citenamefont {Abbott}\ \emph
  {et~al.}(2021{\natexlab{b}})\citenamefont {Abbott} \emph
  {et~al.}}]{LIGOScientific:2021djp}%
  \BibitemOpen
  \bibfield  {author} {\bibinfo {author} {\bibfnamefont {R.}~\bibnamefont
  {Abbott}} \emph {et~al.} (\bibinfo {collaboration} {LIGO Scientific, VIRGO,
  KAGRA}),\ }\bibfield  {title} {\enquote {\bibinfo {title} {Gwtc-3: Compact
  binary coalescences observed by ligo and virgo during the second part of the
  third observing run},}\ }\href@noop {} {\  (\bibinfo {year}
  {2021}{\natexlab{b}})},\ \Eprint {http://arxiv.org/abs/2111.03606}
  {arXiv:2111.03606 [gr-qc]} \BibitemShut {NoStop}%
\bibitem [{\citenamefont {Abbott}\ \emph
  {et~al.}(2019{\natexlab{b}})\citenamefont {Abbott} \emph
  {et~al.}}]{LIGOScientific:2019fpa}%
  \BibitemOpen
  \bibfield  {author} {\bibinfo {author} {\bibfnamefont {B.~P.}\ \bibnamefont
  {Abbott}} \emph {et~al.} (\bibinfo {collaboration} {LIGO Scientific,
  Virgo}),\ }\bibfield  {title} {\enquote {\bibinfo {title} {{Tests of General
  Relativity with the Binary Black Hole Signals from the LIGO-Virgo Catalog
  GWTC-1}},}\ }\href {\doibase 10.1103/PhysRevD.100.104036} {\bibfield
  {journal} {\bibinfo  {journal} {Phys. Rev. D}\ }\textbf {\bibinfo {volume}
  {100}},\ \bibinfo {pages} {104036} (\bibinfo {year} {2019}{\natexlab{b}})},\
  \Eprint {http://arxiv.org/abs/1903.04467} {arXiv:1903.04467 [gr-qc]}
  \BibitemShut {NoStop}%
\bibitem [{\citenamefont {Abbott}\ \emph
  {et~al.}(2021{\natexlab{c}})\citenamefont {Abbott} \emph
  {et~al.}}]{LIGOScientific:2020tif}%
  \BibitemOpen
  \bibfield  {author} {\bibinfo {author} {\bibfnamefont {R.}~\bibnamefont
  {Abbott}} \emph {et~al.} (\bibinfo {collaboration} {LIGO Scientific,
  Virgo}),\ }\bibfield  {title} {\enquote {\bibinfo {title} {{Tests of general
  relativity with binary black holes from the second LIGO-Virgo
  gravitational-wave transient catalog}},}\ }\href {\doibase
  10.1103/PhysRevD.103.122002} {\bibfield  {journal} {\bibinfo  {journal}
  {Phys. Rev. D}\ }\textbf {\bibinfo {volume} {103}},\ \bibinfo {pages}
  {122002} (\bibinfo {year} {2021}{\natexlab{c}})},\ \Eprint
  {http://arxiv.org/abs/2010.14529} {arXiv:2010.14529 [gr-qc]} \BibitemShut
  {NoStop}%
\bibitem [{\citenamefont {Abbott}\ \emph
  {et~al.}(2021{\natexlab{d}})\citenamefont {Abbott} \emph
  {et~al.}}]{LIGOScientific:2021sio}%
  \BibitemOpen
  \bibfield  {author} {\bibinfo {author} {\bibfnamefont {R.}~\bibnamefont
  {Abbott}} \emph {et~al.} (\bibinfo {collaboration} {LIGO Scientific, VIRGO,
  KAGRA}),\ }\bibfield  {title} {\enquote {\bibinfo {title} {{Tests of General
  Relativity with GWTC-3}},}\ }\href@noop {} {\  (\bibinfo {year}
  {2021}{\natexlab{d}})},\ \Eprint {http://arxiv.org/abs/2112.06861}
  {arXiv:2112.06861 [gr-qc]} \BibitemShut {NoStop}%
\bibitem [{\citenamefont {Staelens}(2019)}]{Staelens:2019gzt}%
  \BibitemOpen
  \bibfield  {author} {\bibinfo {author} {\bibfnamefont {Michael}\ \bibnamefont
  {Staelens}} (\bibinfo {collaboration} {MoEDAL}),\ }\bibfield  {title}
  {\enquote {\bibinfo {title} {{Recent Results and Future Plans of the MoEDAL
  Experiment}},}\ }in\ \href@noop {} {\emph {\bibinfo {booktitle} {{Meeting of
  the Division of Particles and Fields of the American Physical Society}}}}\
  (\bibinfo {year} {2019})\ \Eprint {http://arxiv.org/abs/1910.05772}
  {arXiv:1910.05772 [hep-ex]} \BibitemShut {NoStop}%
\bibitem [{\citenamefont {Kobayashi}(2021)}]{Kobayashi:2021des}%
  \BibitemOpen
  \bibfield  {author} {\bibinfo {author} {\bibfnamefont {Takeshi}\ \bibnamefont
  {Kobayashi}},\ }\bibfield  {title} {\enquote {\bibinfo {title}
  {{Monopole-antimonopole pair production in primordial magnetic fields}},}\
  }\href {\doibase 10.1103/PhysRevD.104.043501} {\bibfield  {journal} {\bibinfo
   {journal} {Phys. Rev. D}\ }\textbf {\bibinfo {volume} {104}},\ \bibinfo
  {pages} {043501} (\bibinfo {year} {2021})},\ \Eprint
  {http://arxiv.org/abs/2105.12776} {arXiv:2105.12776 [hep-ph]} \BibitemShut
  {NoStop}%
\bibitem [{\citenamefont {Maldacena}(2021)}]{Maldacena:2020skw}%
  \BibitemOpen
  \bibfield  {author} {\bibinfo {author} {\bibfnamefont {Juan}\ \bibnamefont
  {Maldacena}},\ }\bibfield  {title} {\enquote {\bibinfo {title} {{Comments on
  magnetic black holes}},}\ }\href {\doibase 10.1007/JHEP04(2021)079}
  {\bibfield  {journal} {\bibinfo  {journal} {JHEP}\ }\textbf {\bibinfo
  {volume} {04}},\ \bibinfo {pages} {079} (\bibinfo {year} {2021})},\ \Eprint
  {http://arxiv.org/abs/2004.06084} {arXiv:2004.06084 [hep-th]} \BibitemShut
  {NoStop}%
\bibitem [{\citenamefont {Bai}\ \emph {et~al.}(2020)\citenamefont {Bai},
  \citenamefont {Berger}, \citenamefont {Korwar},\ and\ \citenamefont
  {Orlofsky}}]{Bai:2020spd}%
  \BibitemOpen
  \bibfield  {author} {\bibinfo {author} {\bibfnamefont {Yang}\ \bibnamefont
  {Bai}}, \bibinfo {author} {\bibfnamefont {Joshua}\ \bibnamefont {Berger}},
  \bibinfo {author} {\bibfnamefont {Mrunal}\ \bibnamefont {Korwar}}, \ and\
  \bibinfo {author} {\bibfnamefont {Nicholas}\ \bibnamefont {Orlofsky}},\
  }\bibfield  {title} {\enquote {\bibinfo {title} {{Phenomenology of magnetic
  black holes with electroweak-symmetric coronas}},}\ }\href {\doibase
  10.1007/JHEP10(2020)210} {\bibfield  {journal} {\bibinfo  {journal} {JHEP}\
  }\textbf {\bibinfo {volume} {10}},\ \bibinfo {pages} {210} (\bibinfo {year}
  {2020})},\ \Eprint {http://arxiv.org/abs/2007.03703} {arXiv:2007.03703
  [hep-ph]} \BibitemShut {NoStop}%
\bibitem [{\citenamefont {Liu}\ \emph {et~al.}(2020{\natexlab{a}})\citenamefont
  {Liu}, \citenamefont {Christiansen}, \citenamefont {Guo}, \citenamefont
  {Cai},\ and\ \citenamefont {Kim}}]{Liu:2020vsy}%
  \BibitemOpen
  \bibfield  {author} {\bibinfo {author} {\bibfnamefont {Lang}\ \bibnamefont
  {Liu}}, \bibinfo {author} {\bibfnamefont {\O{}yvind}\ \bibnamefont
  {Christiansen}}, \bibinfo {author} {\bibfnamefont {Zong-Kuan}\ \bibnamefont
  {Guo}}, \bibinfo {author} {\bibfnamefont {Rong-Gen}\ \bibnamefont {Cai}}, \
  and\ \bibinfo {author} {\bibfnamefont {Sang~Pyo}\ \bibnamefont {Kim}},\
  }\bibfield  {title} {\enquote {\bibinfo {title} {{Gravitational and
  electromagnetic radiation from binary black holes with electric and magnetic
  charges: Circular orbits on a cone}},}\ }\href {\doibase
  10.1103/PhysRevD.102.103520} {\bibfield  {journal} {\bibinfo  {journal}
  {Phys. Rev. D}\ }\textbf {\bibinfo {volume} {102}},\ \bibinfo {pages}
  {103520} (\bibinfo {year} {2020}{\natexlab{a}})},\ \Eprint
  {http://arxiv.org/abs/2008.02326} {arXiv:2008.02326 [gr-qc]} \BibitemShut
  {NoStop}%
\bibitem [{\citenamefont {Ghosh}\ \emph {et~al.}(2021)\citenamefont {Ghosh},
  \citenamefont {Thalapillil},\ and\ \citenamefont {Ullah}}]{Ghosh:2020tdu}%
  \BibitemOpen
  \bibfield  {author} {\bibinfo {author} {\bibfnamefont {Diptimoy}\
  \bibnamefont {Ghosh}}, \bibinfo {author} {\bibfnamefont {Arun}\ \bibnamefont
  {Thalapillil}}, \ and\ \bibinfo {author} {\bibfnamefont {Farman}\
  \bibnamefont {Ullah}},\ }\bibfield  {title} {\enquote {\bibinfo {title}
  {{Astrophysical hints for magnetic black holes}},}\ }\href {\doibase
  10.1103/PhysRevD.103.023006} {\bibfield  {journal} {\bibinfo  {journal}
  {Phys. Rev. D}\ }\textbf {\bibinfo {volume} {103}},\ \bibinfo {pages}
  {023006} (\bibinfo {year} {2021})},\ \Eprint
  {http://arxiv.org/abs/2009.03363} {arXiv:2009.03363 [hep-ph]} \BibitemShut
  {NoStop}%
\bibitem [{\citenamefont {Liu}\ \emph {et~al.}(2021)\citenamefont {Liu},
  \citenamefont {Christiansen}, \citenamefont {Ruan}, \citenamefont {Guo},
  \citenamefont {Cai},\ and\ \citenamefont {Kim}}]{Liu:2020bag}%
  \BibitemOpen
  \bibfield  {author} {\bibinfo {author} {\bibfnamefont {Lang}\ \bibnamefont
  {Liu}}, \bibinfo {author} {\bibfnamefont {\O{}yvind}\ \bibnamefont
  {Christiansen}}, \bibinfo {author} {\bibfnamefont {Wen-Hong}\ \bibnamefont
  {Ruan}}, \bibinfo {author} {\bibfnamefont {Zong-Kuan}\ \bibnamefont {Guo}},
  \bibinfo {author} {\bibfnamefont {Rong-Gen}\ \bibnamefont {Cai}}, \ and\
  \bibinfo {author} {\bibfnamefont {Sang~Pyo}\ \bibnamefont {Kim}},\ }\bibfield
   {title} {\enquote {\bibinfo {title} {{Gravitational and electromagnetic
  radiation from binary black holes with electric and magnetic charges:
  elliptical orbits on a cone}},}\ }\href {\doibase
  10.1140/epjc/s10052-021-09849-4} {\bibfield  {journal} {\bibinfo  {journal}
  {Eur. Phys. J. C}\ }\textbf {\bibinfo {volume} {81}},\ \bibinfo {pages}
  {1048} (\bibinfo {year} {2021})},\ \Eprint {http://arxiv.org/abs/2011.13586}
  {arXiv:2011.13586 [gr-qc]} \BibitemShut {NoStop}%
\bibitem [{\citenamefont {Liu}\ \emph {et~al.}(2020{\natexlab{b}})\citenamefont
  {Liu}, \citenamefont {Guo}, \citenamefont {Cai},\ and\ \citenamefont
  {Kim}}]{Liu:2020cds}%
  \BibitemOpen
  \bibfield  {author} {\bibinfo {author} {\bibfnamefont {Lang}\ \bibnamefont
  {Liu}}, \bibinfo {author} {\bibfnamefont {Zong-Kuan}\ \bibnamefont {Guo}},
  \bibinfo {author} {\bibfnamefont {Rong-Gen}\ \bibnamefont {Cai}}, \ and\
  \bibinfo {author} {\bibfnamefont {Sang~Pyo}\ \bibnamefont {Kim}},\ }\bibfield
   {title} {\enquote {\bibinfo {title} {{Merger rate distribution of primordial
  black hole binaries with electric charges}},}\ }\href {\doibase
  10.1103/PhysRevD.102.043508} {\bibfield  {journal} {\bibinfo  {journal}
  {Phys. Rev. D}\ }\textbf {\bibinfo {volume} {102}},\ \bibinfo {pages}
  {043508} (\bibinfo {year} {2020}{\natexlab{b}})},\ \Eprint
  {http://arxiv.org/abs/2001.02984} {arXiv:2001.02984 [astro-ph.CO]}
  \BibitemShut {NoStop}%
\bibitem [{\citenamefont {Zilhao}\ \emph {et~al.}(2012)\citenamefont {Zilhao},
  \citenamefont {Cardoso}, \citenamefont {Herdeiro}, \citenamefont {Lehner},\
  and\ \citenamefont {Sperhake}}]{Zilhao:2012gp}%
  \BibitemOpen
  \bibfield  {author} {\bibinfo {author} {\bibfnamefont {Miguel}\ \bibnamefont
  {Zilhao}}, \bibinfo {author} {\bibfnamefont {Vitor}\ \bibnamefont {Cardoso}},
  \bibinfo {author} {\bibfnamefont {Carlos}\ \bibnamefont {Herdeiro}}, \bibinfo
  {author} {\bibfnamefont {Luis}\ \bibnamefont {Lehner}}, \ and\ \bibinfo
  {author} {\bibfnamefont {Ulrich}\ \bibnamefont {Sperhake}},\ }\bibfield
  {title} {\enquote {\bibinfo {title} {{Collisions of charged black holes}},}\
  }\href {\doibase 10.1103/PhysRevD.85.124062} {\bibfield  {journal} {\bibinfo
  {journal} {Phys. Rev. D}\ }\textbf {\bibinfo {volume} {85}},\ \bibinfo
  {pages} {124062} (\bibinfo {year} {2012})},\ \Eprint
  {http://arxiv.org/abs/1205.1063} {arXiv:1205.1063 [gr-qc]} \BibitemShut
  {NoStop}%
\bibitem [{\citenamefont {Zilh\~ao}\ \emph {et~al.}(2014)\citenamefont
  {Zilh\~ao}, \citenamefont {Cardoso}, \citenamefont {Herdeiro}, \citenamefont
  {Lehner},\ and\ \citenamefont {Sperhake}}]{Zilhao:2013nda}%
  \BibitemOpen
  \bibfield  {author} {\bibinfo {author} {\bibfnamefont {Miguel}\ \bibnamefont
  {Zilh\~ao}}, \bibinfo {author} {\bibfnamefont {Vitor}\ \bibnamefont
  {Cardoso}}, \bibinfo {author} {\bibfnamefont {Carlos}\ \bibnamefont
  {Herdeiro}}, \bibinfo {author} {\bibfnamefont {Luis}\ \bibnamefont {Lehner}},
  \ and\ \bibinfo {author} {\bibfnamefont {Ulrich}\ \bibnamefont {Sperhake}},\
  }\bibfield  {title} {\enquote {\bibinfo {title} {{Collisions of oppositely
  charged black holes}},}\ }\href {\doibase 10.1103/PhysRevD.89.044008}
  {\bibfield  {journal} {\bibinfo  {journal} {Phys. Rev. D}\ }\textbf {\bibinfo
  {volume} {89}},\ \bibinfo {pages} {044008} (\bibinfo {year} {2014})},\
  \Eprint {http://arxiv.org/abs/1311.6483} {arXiv:1311.6483 [gr-qc]}
  \BibitemShut {NoStop}%
\bibitem [{\citenamefont {Liebling}\ and\ \citenamefont
  {Palenzuela}(2016)}]{Liebling:2016orx}%
  \BibitemOpen
  \bibfield  {author} {\bibinfo {author} {\bibfnamefont {Steven~L.}\
  \bibnamefont {Liebling}}\ and\ \bibinfo {author} {\bibfnamefont {Carlos}\
  \bibnamefont {Palenzuela}},\ }\bibfield  {title} {\enquote {\bibinfo {title}
  {{Electromagnetic Luminosity of the Coalescence of Charged Black Hole
  Binaries}},}\ }\href {\doibase 10.1103/PhysRevD.94.064046} {\bibfield
  {journal} {\bibinfo  {journal} {Phys. Rev.}\ }\textbf {\bibinfo {volume}
  {D94}},\ \bibinfo {pages} {064046} (\bibinfo {year} {2016})},\ \Eprint
  {http://arxiv.org/abs/1607.02140} {arXiv:1607.02140 [gr-qc]} \BibitemShut
  {NoStop}%
%%CITATION = ARXIV:1607.02140;%%
\bibitem [{\citenamefont {Toshmatov}\ \emph {et~al.}(2018)\citenamefont
  {Toshmatov}, \citenamefont {Stuchl\'\i{}k}, \citenamefont {Schee},\ and\
  \citenamefont {Ahmedov}}]{Toshmatov:2018tyo}%
  \BibitemOpen
  \bibfield  {author} {\bibinfo {author} {\bibfnamefont {Bobir}\ \bibnamefont
  {Toshmatov}}, \bibinfo {author} {\bibfnamefont {Zden\v{e}k}\ \bibnamefont
  {Stuchl\'\i{}k}}, \bibinfo {author} {\bibfnamefont {Jan}\ \bibnamefont
  {Schee}}, \ and\ \bibinfo {author} {\bibfnamefont {Bobomurat}\ \bibnamefont
  {Ahmedov}},\ }\bibfield  {title} {\enquote {\bibinfo {title}
  {{Electromagnetic perturbations of black holes in general relativity coupled
  to nonlinear electrodynamics}},}\ }\href {\doibase
  10.1103/PhysRevD.97.084058} {\bibfield  {journal} {\bibinfo  {journal} {Phys.
  Rev. D}\ }\textbf {\bibinfo {volume} {97}},\ \bibinfo {pages} {084058}
  (\bibinfo {year} {2018})},\ \Eprint {http://arxiv.org/abs/1805.00240}
  {arXiv:1805.00240 [gr-qc]} \BibitemShut {NoStop}%
\bibitem [{\citenamefont {Bai}\ and\ \citenamefont
  {Orlofsky}(2020)}]{Bai:2019zcd}%
  \BibitemOpen
  \bibfield  {author} {\bibinfo {author} {\bibfnamefont {Yang}\ \bibnamefont
  {Bai}}\ and\ \bibinfo {author} {\bibfnamefont {Nicholas}\ \bibnamefont
  {Orlofsky}},\ }\bibfield  {title} {\enquote {\bibinfo {title} {{Primordial
  Extremal Black Holes as Dark Matter}},}\ }\href {\doibase
  10.1103/PhysRevD.101.055006} {\bibfield  {journal} {\bibinfo  {journal}
  {Phys. Rev. D}\ }\textbf {\bibinfo {volume} {101}},\ \bibinfo {pages}
  {055006} (\bibinfo {year} {2020})},\ \Eprint
  {http://arxiv.org/abs/1906.04858} {arXiv:1906.04858 [hep-ph]} \BibitemShut
  {NoStop}%
\bibitem [{\citenamefont {Allahyari}\ \emph {et~al.}(2020)\citenamefont
  {Allahyari}, \citenamefont {Khodadi}, \citenamefont {Vagnozzi},\ and\
  \citenamefont {Mota}}]{Allahyari:2019jqz}%
  \BibitemOpen
  \bibfield  {author} {\bibinfo {author} {\bibfnamefont {Alireza}\ \bibnamefont
  {Allahyari}}, \bibinfo {author} {\bibfnamefont {Mohsen}\ \bibnamefont
  {Khodadi}}, \bibinfo {author} {\bibfnamefont {Sunny}\ \bibnamefont
  {Vagnozzi}}, \ and\ \bibinfo {author} {\bibfnamefont {David~F.}\ \bibnamefont
  {Mota}},\ }\bibfield  {title} {\enquote {\bibinfo {title} {{Magnetically
  charged black holes from non-linear electrodynamics and the Event Horizon
  Telescope}},}\ }\href {\doibase 10.1088/1475-7516/2020/02/003} {\bibfield
  {journal} {\bibinfo  {journal} {JCAP}\ }\textbf {\bibinfo {volume} {02}},\
  \bibinfo {pages} {003} (\bibinfo {year} {2020})},\ \Eprint
  {http://arxiv.org/abs/1912.08231} {arXiv:1912.08231 [gr-qc]} \BibitemShut
  {NoStop}%
\bibitem [{\citenamefont {Christiansen}\ \emph {et~al.}(2021)\citenamefont
  {Christiansen}, \citenamefont {Beltr\'an~Jim\'enez},\ and\ \citenamefont
  {Mota}}]{Christiansen:2020pnv}%
  \BibitemOpen
  \bibfield  {author} {\bibinfo {author} {\bibfnamefont {\O{}yvind}\
  \bibnamefont {Christiansen}}, \bibinfo {author} {\bibfnamefont {Jose}\
  \bibnamefont {Beltr\'an~Jim\'enez}}, \ and\ \bibinfo {author} {\bibfnamefont
  {David~F.}\ \bibnamefont {Mota}},\ }\bibfield  {title} {\enquote {\bibinfo
  {title} {{Charged Black Hole Mergers: Orbit Circularisation and Chirp Mass
  Bias}},}\ }\href {\doibase 10.1088/1361-6382/abdaf5} {\bibfield  {journal}
  {\bibinfo  {journal} {Class. Quant. Grav.}\ }\textbf {\bibinfo {volume}
  {38}},\ \bibinfo {pages} {075017} (\bibinfo {year} {2021})},\ \Eprint
  {http://arxiv.org/abs/2003.11452} {arXiv:2003.11452 [gr-qc]} \BibitemShut
  {NoStop}%
\bibitem [{\citenamefont {Wang}\ \emph {et~al.}(2021)\citenamefont {Wang},
  \citenamefont {Li}, \citenamefont {Jiang}, \citenamefont {Yuan},
  \citenamefont {Hu},\ and\ \citenamefont {Fan}}]{Wang_2021}%
  \BibitemOpen
  \bibfield  {author} {\bibinfo {author} {\bibfnamefont {Hai-Tang}\
  \bibnamefont {Wang}}, \bibinfo {author} {\bibfnamefont {Peng-Cheng}\
  \bibnamefont {Li}}, \bibinfo {author} {\bibfnamefont {Jin-Liang}\
  \bibnamefont {Jiang}}, \bibinfo {author} {\bibfnamefont {Guan-Wen}\
  \bibnamefont {Yuan}}, \bibinfo {author} {\bibfnamefont {Yi-Ming}\
  \bibnamefont {Hu}}, \ and\ \bibinfo {author} {\bibfnamefont {Yi-Zhong}\
  \bibnamefont {Fan}},\ }\bibfield  {title} {\enquote {\bibinfo {title}
  {Constrains on the electric charges of the binary black holes with {GWTC}-1
  events},}\ }\href {\doibase 10.1140/epjc/s10052-021-09555-1} {\bibfield
  {journal} {\bibinfo  {journal} {The European Physical Journal C}\ }\textbf
  {\bibinfo {volume} {81}} (\bibinfo {year} {2021}),\
  10.1140/epjc/s10052-021-09555-1}\BibitemShut {NoStop}%
\bibitem [{\citenamefont {Bozzola}\ and\ \citenamefont
  {Paschalidis}(2021{\natexlab{a}})}]{Bozzola:2020mjx}%
  \BibitemOpen
  \bibfield  {author} {\bibinfo {author} {\bibfnamefont {Gabriele}\
  \bibnamefont {Bozzola}}\ and\ \bibinfo {author} {\bibfnamefont {Vasileios}\
  \bibnamefont {Paschalidis}},\ }\bibfield  {title} {\enquote {\bibinfo {title}
  {{General Relativistic Simulations of the Quasicircular Inspiral and Merger
  of Charged Black Holes: GW150914 and Fundamental Physics Implications}},}\
  }\href {\doibase 10.1103/PhysRevLett.126.041103} {\bibfield  {journal}
  {\bibinfo  {journal} {Phys. Rev. Lett.}\ }\textbf {\bibinfo {volume} {126}},\
  \bibinfo {pages} {041103} (\bibinfo {year} {2021}{\natexlab{a}})},\ \Eprint
  {http://arxiv.org/abs/2006.15764} {arXiv:2006.15764 [gr-qc]} \BibitemShut
  {NoStop}%
\bibitem [{\citenamefont {Kim}\ and\ \citenamefont
  {Kobakhidze}(2020)}]{Kim:2020bhg}%
  \BibitemOpen
  \bibfield  {author} {\bibinfo {author} {\bibfnamefont {Yunho}\ \bibnamefont
  {Kim}}\ and\ \bibinfo {author} {\bibfnamefont {Archil}\ \bibnamefont
  {Kobakhidze}},\ }\bibfield  {title} {\enquote {\bibinfo {title}
  {{Topologically induced black hole charge and its astrophysical
  manifestations}},}\ }\href@noop {} {\  (\bibinfo {year} {2020})},\ \Eprint
  {http://arxiv.org/abs/2008.04506} {arXiv:2008.04506 [gr-qc]} \BibitemShut
  {NoStop}%
\bibitem [{\citenamefont {Cardoso}\ \emph
  {et~al.}(2021{\natexlab{a}})\citenamefont {Cardoso}, \citenamefont {Guo},
  \citenamefont {Macedo},\ and\ \citenamefont {Pani}}]{Cardoso:2020nst}%
  \BibitemOpen
  \bibfield  {author} {\bibinfo {author} {\bibfnamefont {Vitor}\ \bibnamefont
  {Cardoso}}, \bibinfo {author} {\bibfnamefont {Wen-Di}\ \bibnamefont {Guo}},
  \bibinfo {author} {\bibfnamefont {Caio F.~B.}\ \bibnamefont {Macedo}}, \ and\
  \bibinfo {author} {\bibfnamefont {Paolo}\ \bibnamefont {Pani}},\ }\bibfield
  {title} {\enquote {\bibinfo {title} {{The tune of the Universe: the role of
  plasma in tests of strong-field gravity}},}\ }\href {\doibase
  10.1093/mnras/stab404} {\bibfield  {journal} {\bibinfo  {journal} {Mon. Not.
  Roy. Astron. Soc.}\ }\textbf {\bibinfo {volume} {503}},\ \bibinfo {pages}
  {563--573} (\bibinfo {year} {2021}{\natexlab{a}})},\ \Eprint
  {http://arxiv.org/abs/2009.07287} {arXiv:2009.07287 [gr-qc]} \BibitemShut
  {NoStop}%
\bibitem [{\citenamefont {McInnes}(2021{\natexlab{a}})}]{McInnes:2020gxx}%
  \BibitemOpen
  \bibfield  {author} {\bibinfo {author} {\bibfnamefont {Brett}\ \bibnamefont
  {McInnes}},\ }\bibfield  {title} {\enquote {\bibinfo {title} {{About Magnetic
  AdS Black Holes}},}\ }\href {\doibase 10.1007/JHEP03(2021)068} {\bibfield
  {journal} {\bibinfo  {journal} {JHEP}\ }\textbf {\bibinfo {volume} {03}},\
  \bibinfo {pages} {068} (\bibinfo {year} {2021}{\natexlab{a}})},\ \Eprint
  {http://arxiv.org/abs/2011.07700} {arXiv:2011.07700 [gr-qc]} \BibitemShut
  {NoStop}%
\bibitem [{\citenamefont {Bai}\ and\ \citenamefont
  {Korwar}(2021)}]{Bai:2020ezy}%
  \BibitemOpen
  \bibfield  {author} {\bibinfo {author} {\bibfnamefont {Yang}\ \bibnamefont
  {Bai}}\ and\ \bibinfo {author} {\bibfnamefont {Mrunal}\ \bibnamefont
  {Korwar}},\ }\bibfield  {title} {\enquote {\bibinfo {title} {{Hairy Magnetic
  and Dyonic Black Holes in the Standard Model}},}\ }\href {\doibase
  10.1007/JHEP04(2021)119} {\bibfield  {journal} {\bibinfo  {journal} {JHEP}\
  }\textbf {\bibinfo {volume} {04}},\ \bibinfo {pages} {119} (\bibinfo {year}
  {2021})},\ \Eprint {http://arxiv.org/abs/2012.15430} {arXiv:2012.15430
  [hep-ph]} \BibitemShut {NoStop}%
\bibitem [{\citenamefont {Diamond}\ and\ \citenamefont
  {Kaplan}(2022)}]{Diamond:2021scl}%
  \BibitemOpen
  \bibfield  {author} {\bibinfo {author} {\bibfnamefont {Melissa~D.}\
  \bibnamefont {Diamond}}\ and\ \bibinfo {author} {\bibfnamefont {David~E.}\
  \bibnamefont {Kaplan}},\ }\bibfield  {title} {\enquote {\bibinfo {title}
  {{Constraints on relic magnetic black holes}},}\ }\href {\doibase
  10.1007/JHEP03(2022)157} {\bibfield  {journal} {\bibinfo  {journal} {JHEP}\
  }\textbf {\bibinfo {volume} {03}},\ \bibinfo {pages} {157} (\bibinfo {year}
  {2022})},\ \Eprint {http://arxiv.org/abs/2103.01850} {arXiv:2103.01850
  [hep-ph]} \BibitemShut {NoStop}%
\bibitem [{\citenamefont {Bozzola}\ and\ \citenamefont
  {Paschalidis}(2021{\natexlab{b}})}]{Bozzola:2021elc}%
  \BibitemOpen
  \bibfield  {author} {\bibinfo {author} {\bibfnamefont {Gabriele}\
  \bibnamefont {Bozzola}}\ and\ \bibinfo {author} {\bibfnamefont {Vasileios}\
  \bibnamefont {Paschalidis}},\ }\bibfield  {title} {\enquote {\bibinfo {title}
  {{Numerical-relativity simulations of the quasicircular inspiral and merger
  of nonspinning, charged black holes: Methods and comparison with approximate
  approaches}},}\ }\href {\doibase 10.1103/PhysRevD.104.044004} {\bibfield
  {journal} {\bibinfo  {journal} {Phys. Rev. D}\ }\textbf {\bibinfo {volume}
  {104}},\ \bibinfo {pages} {044004} (\bibinfo {year} {2021}{\natexlab{b}})},\
  \Eprint {http://arxiv.org/abs/2104.06978} {arXiv:2104.06978 [gr-qc]}
  \BibitemShut {NoStop}%
\bibitem [{\citenamefont {McInnes}(2021{\natexlab{b}})}]{McInnes:2021frb}%
  \BibitemOpen
  \bibfield  {author} {\bibinfo {author} {\bibfnamefont {Brett}\ \bibnamefont
  {McInnes}},\ }\bibfield  {title} {\enquote {\bibinfo {title} {{The weak
  gravity conjecture requires the existence of exotic AdS black holes}},}\
  }\href {\doibase 10.1016/j.nuclphysb.2021.115525} {\bibfield  {journal}
  {\bibinfo  {journal} {Nucl. Phys. B}\ }\textbf {\bibinfo {volume} {971}},\
  \bibinfo {pages} {115525} (\bibinfo {year} {2021}{\natexlab{b}})},\ \Eprint
  {http://arxiv.org/abs/2104.07373} {arXiv:2104.07373 [gr-qc]} \BibitemShut
  {NoStop}%
\bibitem [{\citenamefont {Kritos}\ and\ \citenamefont
  {Silk}(2022)}]{Kritos:2021nsf}%
  \BibitemOpen
  \bibfield  {author} {\bibinfo {author} {\bibfnamefont {Konstantinos}\
  \bibnamefont {Kritos}}\ and\ \bibinfo {author} {\bibfnamefont {Joseph}\
  \bibnamefont {Silk}},\ }\bibfield  {title} {\enquote {\bibinfo {title}
  {{Mergers of maximally charged primordial black holes}},}\ }\href {\doibase
  10.1103/PhysRevD.105.063011} {\bibfield  {journal} {\bibinfo  {journal}
  {Phys. Rev. D}\ }\textbf {\bibinfo {volume} {105}},\ \bibinfo {pages}
  {063011} (\bibinfo {year} {2022})},\ \Eprint
  {http://arxiv.org/abs/2109.09769} {arXiv:2109.09769 [gr-qc]} \BibitemShut
  {NoStop}%
\bibitem [{\citenamefont {Hou}\ \emph {et~al.}(2022)\citenamefont {Hou},
  \citenamefont {Tian}, \citenamefont {Cao},\ and\ \citenamefont
  {Zhu}}]{Hou:2021suj}%
  \BibitemOpen
  \bibfield  {author} {\bibinfo {author} {\bibfnamefont {Shaoqi}\ \bibnamefont
  {Hou}}, \bibinfo {author} {\bibfnamefont {Shuxun}\ \bibnamefont {Tian}},
  \bibinfo {author} {\bibfnamefont {Shuo}\ \bibnamefont {Cao}}, \ and\ \bibinfo
  {author} {\bibfnamefont {Zong-Hong}\ \bibnamefont {Zhu}},\ }\bibfield
  {title} {\enquote {\bibinfo {title} {{Dark photon bursts from compact binary
  systems and constraints}},}\ }\href {\doibase 10.1103/PhysRevD.105.064022}
  {\bibfield  {journal} {\bibinfo  {journal} {Phys. Rev. D}\ }\textbf {\bibinfo
  {volume} {105}},\ \bibinfo {pages} {064022} (\bibinfo {year} {2022})},\
  \Eprint {http://arxiv.org/abs/2110.05084} {arXiv:2110.05084 [hep-ph]}
  \BibitemShut {NoStop}%
\bibitem [{\citenamefont {Benavides-Gallego}\ and\ \citenamefont
  {Han}(2021)}]{Benavides-Gallego:2021the}%
  \BibitemOpen
  \bibfield  {author} {\bibinfo {author} {\bibfnamefont {Carlos~A.}\
  \bibnamefont {Benavides-Gallego}}\ and\ \bibinfo {author} {\bibfnamefont
  {Wen-Biao}\ \bibnamefont {Han}},\ }\bibfield  {title} {\enquote {\bibinfo
  {title} {{Phenomenological model for the electromagnetic response of a black
  hole binary immersed in magnetic field}},}\ }\href@noop {} {\  (\bibinfo
  {year} {2021})},\ \Eprint {http://arxiv.org/abs/2111.04323} {arXiv:2111.04323
  [gr-qc]} \BibitemShut {NoStop}%
\bibitem [{\citenamefont {Diamond}\ \emph {et~al.}(2021)\citenamefont
  {Diamond}, \citenamefont {Kaplan},\ and\ \citenamefont
  {Rajendran}}]{Diamond:2021dth}%
  \BibitemOpen
  \bibfield  {author} {\bibinfo {author} {\bibfnamefont {Melissa~D.}\
  \bibnamefont {Diamond}}, \bibinfo {author} {\bibfnamefont {David~E.}\
  \bibnamefont {Kaplan}}, \ and\ \bibinfo {author} {\bibfnamefont {Surjeet}\
  \bibnamefont {Rajendran}},\ }\bibfield  {title} {\enquote {\bibinfo {title}
  {{Binary Collisions of Dark Matter Blobs}},}\ }\href@noop {} {\  (\bibinfo
  {year} {2021})},\ \Eprint {http://arxiv.org/abs/2112.09147} {arXiv:2112.09147
  [hep-ph]} \BibitemShut {NoStop}%
\bibitem [{\citenamefont {Ackerman}\ \emph {et~al.}(2009)\citenamefont
  {Ackerman}, \citenamefont {Buckley}, \citenamefont {Carroll},\ and\
  \citenamefont {Kamionkowski}}]{Ackerman:2008kmp}%
  \BibitemOpen
  \bibfield  {author} {\bibinfo {author} {\bibfnamefont {Lotty}\ \bibnamefont
  {Ackerman}}, \bibinfo {author} {\bibfnamefont {Matthew~R.}\ \bibnamefont
  {Buckley}}, \bibinfo {author} {\bibfnamefont {Sean~M.}\ \bibnamefont
  {Carroll}}, \ and\ \bibinfo {author} {\bibfnamefont {Marc}\ \bibnamefont
  {Kamionkowski}},\ }\bibfield  {title} {\enquote {\bibinfo {title} {{Dark
  Matter and Dark Radiation}},}\ }\href {\doibase 10.1103/PhysRevD.79.023519}
  {\bibfield  {journal} {\bibinfo  {journal} {Phys. Rev. D}\ }\textbf {\bibinfo
  {volume} {79}},\ \bibinfo {pages} {023519} (\bibinfo {year} {2009})},\
  \Eprint {http://arxiv.org/abs/0810.5126} {arXiv:0810.5126 [hep-ph]}
  \BibitemShut {NoStop}%
\bibitem [{\citenamefont {Feng}\ \emph {et~al.}(2009)\citenamefont {Feng},
  \citenamefont {Kaplinghat}, \citenamefont {Tu},\ and\ \citenamefont
  {Yu}}]{Feng:2009mn}%
  \BibitemOpen
  \bibfield  {author} {\bibinfo {author} {\bibfnamefont {Jonathan~L.}\
  \bibnamefont {Feng}}, \bibinfo {author} {\bibfnamefont {Manoj}\ \bibnamefont
  {Kaplinghat}}, \bibinfo {author} {\bibfnamefont {Huitzu}\ \bibnamefont {Tu}},
  \ and\ \bibinfo {author} {\bibfnamefont {Hai-Bo}\ \bibnamefont {Yu}},\
  }\bibfield  {title} {\enquote {\bibinfo {title} {{Hidden Charged Dark
  Matter}},}\ }\href {\doibase 10.1088/1475-7516/2009/07/004} {\bibfield
  {journal} {\bibinfo  {journal} {JCAP}\ }\textbf {\bibinfo {volume} {07}},\
  \bibinfo {pages} {004} (\bibinfo {year} {2009})},\ \Eprint
  {http://arxiv.org/abs/0905.3039} {arXiv:0905.3039 [hep-ph]} \BibitemShut
  {NoStop}%
\bibitem [{\citenamefont {Foot}\ and\ \citenamefont
  {Vagnozzi}(2015{\natexlab{a}})}]{Foot:2014uba}%
  \BibitemOpen
  \bibfield  {author} {\bibinfo {author} {\bibfnamefont {R.}~\bibnamefont
  {Foot}}\ and\ \bibinfo {author} {\bibfnamefont {S.}~\bibnamefont
  {Vagnozzi}},\ }\bibfield  {title} {\enquote {\bibinfo {title} {{Dissipative
  hidden sector dark matter}},}\ }\href {\doibase 10.1103/PhysRevD.91.023512}
  {\bibfield  {journal} {\bibinfo  {journal} {Phys. Rev. D}\ }\textbf {\bibinfo
  {volume} {91}},\ \bibinfo {pages} {023512} (\bibinfo {year}
  {2015}{\natexlab{a}})},\ \Eprint {http://arxiv.org/abs/1409.7174}
  {arXiv:1409.7174 [hep-ph]} \BibitemShut {NoStop}%
\bibitem [{\citenamefont {Foot}\ and\ \citenamefont
  {Vagnozzi}(2015{\natexlab{b}})}]{Foot:2014osa}%
  \BibitemOpen
  \bibfield  {author} {\bibinfo {author} {\bibfnamefont {R.}~\bibnamefont
  {Foot}}\ and\ \bibinfo {author} {\bibfnamefont {S.}~\bibnamefont
  {Vagnozzi}},\ }\bibfield  {title} {\enquote {\bibinfo {title} {{Diurnal
  modulation signal from dissipative hidden sector dark matter}},}\ }\href
  {\doibase 10.1016/j.physletb.2015.06.063} {\bibfield  {journal} {\bibinfo
  {journal} {Phys. Lett. B}\ }\textbf {\bibinfo {volume} {748}},\ \bibinfo
  {pages} {61--66} (\bibinfo {year} {2015}{\natexlab{b}})},\ \Eprint
  {http://arxiv.org/abs/1412.0762} {arXiv:1412.0762 [hep-ph]} \BibitemShut
  {NoStop}%
\bibitem [{\citenamefont {Moffat}(2006)}]{Moffat:2005si}%
  \BibitemOpen
  \bibfield  {author} {\bibinfo {author} {\bibfnamefont {J.~W.}\ \bibnamefont
  {Moffat}},\ }\bibfield  {title} {\enquote {\bibinfo {title}
  {{Scalar-tensor-vector gravity theory}},}\ }\href {\doibase
  10.1088/1475-7516/2006/03/004} {\bibfield  {journal} {\bibinfo  {journal}
  {JCAP}\ }\textbf {\bibinfo {volume} {03}},\ \bibinfo {pages} {004} (\bibinfo
  {year} {2006})},\ \Eprint {http://arxiv.org/abs/gr-qc/0506021}
  {arXiv:gr-qc/0506021} \BibitemShut {NoStop}%
\bibitem [{\citenamefont {Cardoso}\ \emph {et~al.}(2016)\citenamefont
  {Cardoso}, \citenamefont {Macedo}, \citenamefont {Pani},\ and\ \citenamefont
  {Ferrari}}]{Cardoso:2016olt}%
  \BibitemOpen
  \bibfield  {author} {\bibinfo {author} {\bibfnamefont {Vitor}\ \bibnamefont
  {Cardoso}}, \bibinfo {author} {\bibfnamefont {Caio F.~B.}\ \bibnamefont
  {Macedo}}, \bibinfo {author} {\bibfnamefont {Paolo}\ \bibnamefont {Pani}}, \
  and\ \bibinfo {author} {\bibfnamefont {Valeria}\ \bibnamefont {Ferrari}},\
  }\bibfield  {title} {\enquote {\bibinfo {title} {{Black holes and
  gravitational waves in models of minicharged dark matter}},}\ }\href
  {\doibase 10.1088/1475-7516/2016/05/054} {\bibfield  {journal} {\bibinfo
  {journal} {JCAP}\ }\textbf {\bibinfo {volume} {05}},\ \bibinfo {pages} {054}
  (\bibinfo {year} {2016})},\ \bibinfo {note} {[Erratum: JCAP 04, E01
  (2020)]},\ \Eprint {http://arxiv.org/abs/1604.07845} {arXiv:1604.07845
  [hep-ph]} \BibitemShut {NoStop}%
\bibitem [{\citenamefont {Cardoso}\ \emph
  {et~al.}(2021{\natexlab{b}})\citenamefont {Cardoso}, \citenamefont {Macedo},\
  and\ \citenamefont {Vicente}}]{Cardoso:2020iji}%
  \BibitemOpen
  \bibfield  {author} {\bibinfo {author} {\bibfnamefont {Vitor}\ \bibnamefont
  {Cardoso}}, \bibinfo {author} {\bibfnamefont {Caio F.~B.}\ \bibnamefont
  {Macedo}}, \ and\ \bibinfo {author} {\bibfnamefont {Rodrigo}\ \bibnamefont
  {Vicente}},\ }\bibfield  {title} {\enquote {\bibinfo {title} {{Eccentricity
  evolution of compact binaries and applications to gravitational-wave
  physics}},}\ }\href {\doibase 10.1103/PhysRevD.103.023015} {\bibfield
  {journal} {\bibinfo  {journal} {Phys. Rev. D}\ }\textbf {\bibinfo {volume}
  {103}},\ \bibinfo {pages} {023015} (\bibinfo {year} {2021}{\natexlab{b}})},\
  \Eprint {http://arxiv.org/abs/2010.15151} {arXiv:2010.15151 [gr-qc]}
  \BibitemShut {NoStop}%
\bibitem [{\citenamefont {Liu}\ and\ \citenamefont
  {Kim}(2022{\natexlab{a}})}]{Liu:2022cuj}%
  \BibitemOpen
  \bibfield  {author} {\bibinfo {author} {\bibfnamefont {Lang}\ \bibnamefont
  {Liu}}\ and\ \bibinfo {author} {\bibfnamefont {Sang~Pyo}\ \bibnamefont
  {Kim}},\ }\bibfield  {title} {\enquote {\bibinfo {title} {{Gravitational and
  electromagnetic radiations from binary black holes with electric and magnetic
  charges}},}\ }in\ \href@noop {} {\emph {\bibinfo {booktitle} {{17th
  Italian-Korean Symposium on Relativistic Astrophysics}}}}\ (\bibinfo {year}
  {2022})\ \Eprint {http://arxiv.org/abs/2201.01138} {arXiv:2201.01138 [gr-qc]}
  \BibitemShut {NoStop}%
\bibitem [{\citenamefont {Liu}\ and\ \citenamefont
  {Kim}(2022{\natexlab{b}})}]{Liu:2022wtq}%
  \BibitemOpen
  \bibfield  {author} {\bibinfo {author} {\bibfnamefont {Lang}\ \bibnamefont
  {Liu}}\ and\ \bibinfo {author} {\bibfnamefont {Sang~Pyo}\ \bibnamefont
  {Kim}},\ }\bibfield  {title} {\enquote {\bibinfo {title} {{Merger rate of
  charged black holes from the two-body dynamical capture}},}\ }\href {\doibase
  10.1088/1475-7516/2022/03/059} {\bibfield  {journal} {\bibinfo  {journal}
  {JCAP}\ }\textbf {\bibinfo {volume} {03}},\ \bibinfo {pages} {059} (\bibinfo
  {year} {2022}{\natexlab{b}})},\ \Eprint {http://arxiv.org/abs/2201.02581}
  {arXiv:2201.02581 [gr-qc]} \BibitemShut {NoStop}%
\bibitem [{\citenamefont {Zhang}\ and\ \citenamefont
  {Gong}(2022)}]{Zhang:2022hbt}%
  \BibitemOpen
  \bibfield  {author} {\bibinfo {author} {\bibfnamefont {Chao}\ \bibnamefont
  {Zhang}}\ and\ \bibinfo {author} {\bibfnamefont {Yungui}\ \bibnamefont
  {Gong}},\ }\bibfield  {title} {\enquote {\bibinfo {title} {{Detecting
  electric charge with extreme mass ratio inspirals}},}\ }\href {\doibase
  10.1103/PhysRevD.105.124046} {\bibfield  {journal} {\bibinfo  {journal}
  {Phys. Rev. D}\ }\textbf {\bibinfo {volume} {105}},\ \bibinfo {pages}
  {124046} (\bibinfo {year} {2022})},\ \Eprint
  {http://arxiv.org/abs/2204.08881} {arXiv:2204.08881 [gr-qc]} \BibitemShut
  {NoStop}%
\bibitem [{\citenamefont {Pina}\ \emph {et~al.}(2022)\citenamefont {Pina},
  \citenamefont {Orselli},\ and\ \citenamefont {Pica}}]{Pina:2022dye}%
  \BibitemOpen
  \bibfield  {author} {\bibinfo {author} {\bibfnamefont {D.~Mar\'\i{}n}\
  \bibnamefont {Pina}}, \bibinfo {author} {\bibfnamefont {M.}~\bibnamefont
  {Orselli}}, \ and\ \bibinfo {author} {\bibfnamefont {D.}~\bibnamefont
  {Pica}},\ }\bibfield  {title} {\enquote {\bibinfo {title} {{Event horizon of
  a charged black hole binary merger}},}\ }\href {\doibase
  10.1103/PhysRevD.106.084012} {\bibfield  {journal} {\bibinfo  {journal}
  {Phys. Rev. D}\ }\textbf {\bibinfo {volume} {106}},\ \bibinfo {pages}
  {084012} (\bibinfo {year} {2022})},\ \Eprint
  {http://arxiv.org/abs/2204.08841} {arXiv:2204.08841 [gr-qc]} \BibitemShut
  {NoStop}%
\bibitem [{\citenamefont {Zi}\ \emph {et~al.}(2023)\citenamefont {Zi},
  \citenamefont {Zhou}, \citenamefont {Wang}, \citenamefont {Li}, \citenamefont
  {Zhang},\ and\ \citenamefont {Chen}}]{Zi:2022hcc}%
  \BibitemOpen
  \bibfield  {author} {\bibinfo {author} {\bibfnamefont {Tieguang}\
  \bibnamefont {Zi}}, \bibinfo {author} {\bibfnamefont {Ziqi}\ \bibnamefont
  {Zhou}}, \bibinfo {author} {\bibfnamefont {Hai-Tian}\ \bibnamefont {Wang}},
  \bibinfo {author} {\bibfnamefont {Peng-Cheng}\ \bibnamefont {Li}}, \bibinfo
  {author} {\bibfnamefont {Jian-dong}\ \bibnamefont {Zhang}}, \ and\ \bibinfo
  {author} {\bibfnamefont {Bin}\ \bibnamefont {Chen}},\ }\bibfield  {title}
  {\enquote {\bibinfo {title} {{Analytic kludge waveforms for
  extreme-mass-ratio inspirals of a charged object around a Kerr-Newman black
  hole}},}\ }\href {\doibase 10.1103/PhysRevD.107.023005} {\bibfield  {journal}
  {\bibinfo  {journal} {Phys. Rev. D}\ }\textbf {\bibinfo {volume} {107}},\
  \bibinfo {pages} {023005} (\bibinfo {year} {2023})},\ \Eprint
  {http://arxiv.org/abs/2205.00425} {arXiv:2205.00425 [gr-qc]} \BibitemShut
  {NoStop}%
\bibitem [{\citenamefont {Benavides-Gallego}\ and\ \citenamefont
  {Han}(2023)}]{Benavides-Gallego:2022dpn}%
  \BibitemOpen
  \bibfield  {author} {\bibinfo {author} {\bibfnamefont {Carlos~A.}\
  \bibnamefont {Benavides-Gallego}}\ and\ \bibinfo {author} {\bibfnamefont
  {Wen-Biao}\ \bibnamefont {Han}},\ }\bibfield  {title} {\enquote {\bibinfo
  {title} {{Gravitational waves and electromagnetic radiation from charged
  black hole binaries}},}\ }\href {\doibase 10.3390/sym15020537} {\bibfield
  {journal} {\bibinfo  {journal} {Symmetry}\ }\textbf {\bibinfo {volume}
  {15}},\ \bibinfo {pages} {537} (\bibinfo {year} {2023})},\ \Eprint
  {http://arxiv.org/abs/2209.00874} {arXiv:2209.00874 [gr-qc]} \BibitemShut
  {NoStop}%
\bibitem [{\citenamefont {Estes}\ \emph {et~al.}(2022)\citenamefont {Estes},
  \citenamefont {Kavic}, \citenamefont {Liebling}, \citenamefont {Lippert},\
  and\ \citenamefont {Simonetti}}]{Estes:2022buj}%
  \BibitemOpen
  \bibfield  {author} {\bibinfo {author} {\bibfnamefont {John}\ \bibnamefont
  {Estes}}, \bibinfo {author} {\bibfnamefont {Michael}\ \bibnamefont {Kavic}},
  \bibinfo {author} {\bibfnamefont {Steven~L.}\ \bibnamefont {Liebling}},
  \bibinfo {author} {\bibfnamefont {Matthew}\ \bibnamefont {Lippert}}, \ and\
  \bibinfo {author} {\bibfnamefont {John~H.}\ \bibnamefont {Simonetti}},\
  }\bibfield  {title} {\enquote {\bibinfo {title} {{Stability and Observability
  of Magnetic Primordial Black Hole-Neutron Star Collisions}},}\ }\href@noop {}
  {\  (\bibinfo {year} {2022})},\ \Eprint {http://arxiv.org/abs/2209.06060}
  {arXiv:2209.06060 [astro-ph.HE]} \BibitemShut {NoStop}%
\bibitem [{\citenamefont {Peters}\ and\ \citenamefont
  {Mathews}(1963)}]{Peters:1963ux}%
  \BibitemOpen
  \bibfield  {author} {\bibinfo {author} {\bibfnamefont {P.~C.}\ \bibnamefont
  {Peters}}\ and\ \bibinfo {author} {\bibfnamefont {J.}~\bibnamefont
  {Mathews}},\ }\bibfield  {title} {\enquote {\bibinfo {title} {{Gravitational
  radiation from point masses in a Keplerian orbit}},}\ }\href {\doibase
  10.1103/PhysRev.131.435} {\bibfield  {journal} {\bibinfo  {journal} {Phys.
  Rev.}\ }\textbf {\bibinfo {volume} {131}},\ \bibinfo {pages} {435--439}
  (\bibinfo {year} {1963})}\BibitemShut {NoStop}%
%%CITATION = PHRVA,131,435;%%
\bibitem [{\citenamefont {Zel'dovich}(1967)}]{Zeldovich:1967lct}%
  \BibitemOpen
  \bibfield  {author} {\bibinfo {author} {\bibfnamefont {Ya. B. ;~Novikov}\
  \bibnamefont {Zel'dovich}, \bibfnamefont {I.~D.}},\ }\bibfield  {title}
  {\enquote {\bibinfo {title} {{The Hypothesis of Cores Retarded during
  Expansion and the Hot Cosmological Model}},}\ }\href@noop {} {\bibfield
  {journal} {\bibinfo  {journal} {Soviet Astron. AJ (Engl. Transl. ),}\
  }\textbf {\bibinfo {volume} {10}},\ \bibinfo {pages} {602} (\bibinfo {year}
  {1967})}\BibitemShut {NoStop}%
\bibitem [{\citenamefont {Hawking}(1971)}]{Hawking:1971ei}%
  \BibitemOpen
  \bibfield  {author} {\bibinfo {author} {\bibfnamefont {Stephen}\ \bibnamefont
  {Hawking}},\ }\bibfield  {title} {\enquote {\bibinfo {title}
  {{Gravitationally collapsed objects of very low mass}},}\ }\href@noop {}
  {\bibfield  {journal} {\bibinfo  {journal} {Mon. Not. Roy. Astron. Soc.}\
  }\textbf {\bibinfo {volume} {152}},\ \bibinfo {pages} {75} (\bibinfo {year}
  {1971})}\BibitemShut {NoStop}%
\bibitem [{\citenamefont {Carr}\ and\ \citenamefont
  {Hawking}(1974)}]{Carr:1974nx}%
  \BibitemOpen
  \bibfield  {author} {\bibinfo {author} {\bibfnamefont {Bernard~J.}\
  \bibnamefont {Carr}}\ and\ \bibinfo {author} {\bibfnamefont {S.~W.}\
  \bibnamefont {Hawking}},\ }\bibfield  {title} {\enquote {\bibinfo {title}
  {{Black holes in the early Universe}},}\ }\href@noop {} {\bibfield  {journal}
  {\bibinfo  {journal} {Mon. Not. Roy. Astron. Soc.}\ }\textbf {\bibinfo
  {volume} {168}},\ \bibinfo {pages} {399--415} (\bibinfo {year}
  {1974})}\BibitemShut {NoStop}%
\bibitem [{\citenamefont {Das}\ and\ \citenamefont {Hook}(2021)}]{Das:2021wei}%
  \BibitemOpen
  \bibfield  {author} {\bibinfo {author} {\bibfnamefont {Saurav}\ \bibnamefont
  {Das}}\ and\ \bibinfo {author} {\bibfnamefont {Anson}\ \bibnamefont {Hook}},\
  }\bibfield  {title} {\enquote {\bibinfo {title} {{Black hole production of
  monopoles in the early universe}},}\ }\href {\doibase
  10.1007/JHEP12(2021)145} {\bibfield  {journal} {\bibinfo  {journal} {JHEP}\
  }\textbf {\bibinfo {volume} {12}},\ \bibinfo {pages} {145} (\bibinfo {year}
  {2021})},\ \Eprint {http://arxiv.org/abs/2109.00039} {arXiv:2109.00039
  [hep-ph]} \BibitemShut {NoStop}%
\bibitem [{\citenamefont {Wald}(1974)}]{Wald:1974np}%
  \BibitemOpen
  \bibfield  {author} {\bibinfo {author} {\bibfnamefont {Robert~M.}\
  \bibnamefont {Wald}},\ }\bibfield  {title} {\enquote {\bibinfo {title}
  {{Black hole in a uniform magnetic field}},}\ }\href {\doibase
  10.1103/PhysRevD.10.1680} {\bibfield  {journal} {\bibinfo  {journal} {Phys.
  Rev. D}\ }\textbf {\bibinfo {volume} {10}},\ \bibinfo {pages} {1680--1685}
  (\bibinfo {year} {1974})}\BibitemShut {NoStop}%
\bibitem [{\citenamefont {Sasaki}\ \emph {et~al.}(2018)\citenamefont {Sasaki},
  \citenamefont {Suyama}, \citenamefont {Tanaka},\ and\ \citenamefont
  {Yokoyama}}]{Sasaki:2018dmp}%
  \BibitemOpen
  \bibfield  {author} {\bibinfo {author} {\bibfnamefont {Misao}\ \bibnamefont
  {Sasaki}}, \bibinfo {author} {\bibfnamefont {Teruaki}\ \bibnamefont
  {Suyama}}, \bibinfo {author} {\bibfnamefont {Takahiro}\ \bibnamefont
  {Tanaka}}, \ and\ \bibinfo {author} {\bibfnamefont {Shuichiro}\ \bibnamefont
  {Yokoyama}},\ }\bibfield  {title} {\enquote {\bibinfo {title} {{Primordial
  black holes---perspectives in gravitational wave astronomy}},}\ }\href
  {\doibase 10.1088/1361-6382/aaa7b4} {\bibfield  {journal} {\bibinfo
  {journal} {Class. Quant. Grav.}\ }\textbf {\bibinfo {volume} {35}},\ \bibinfo
  {pages} {063001} (\bibinfo {year} {2018})},\ \Eprint
  {http://arxiv.org/abs/1801.05235} {arXiv:1801.05235 [astro-ph.CO]}
  \BibitemShut {NoStop}%
%%CITATION = ARXIV:1801.05235;%%
\bibitem [{\citenamefont {Carr}\ \emph {et~al.}(2021)\citenamefont {Carr},
  \citenamefont {Kohri}, \citenamefont {Sendouda},\ and\ \citenamefont
  {Yokoyama}}]{Carr:2020gox}%
  \BibitemOpen
  \bibfield  {author} {\bibinfo {author} {\bibfnamefont {Bernard}\ \bibnamefont
  {Carr}}, \bibinfo {author} {\bibfnamefont {Kazunori}\ \bibnamefont {Kohri}},
  \bibinfo {author} {\bibfnamefont {Yuuiti}\ \bibnamefont {Sendouda}}, \ and\
  \bibinfo {author} {\bibfnamefont {Jun'ichi}\ \bibnamefont {Yokoyama}},\
  }\bibfield  {title} {\enquote {\bibinfo {title} {{Constraints on primordial
  black holes}},}\ }\href {\doibase 10.1088/1361-6633/ac1e31} {\bibfield
  {journal} {\bibinfo  {journal} {Rept. Prog. Phys.}\ }\textbf {\bibinfo
  {volume} {84}},\ \bibinfo {pages} {116902} (\bibinfo {year} {2021})},\
  \Eprint {http://arxiv.org/abs/2002.12778} {arXiv:2002.12778 [astro-ph.CO]}
  \BibitemShut {NoStop}%
\bibitem [{\citenamefont {Carr}\ and\ \citenamefont
  {Kuhnel}(2020)}]{Carr:2020xqk}%
  \BibitemOpen
  \bibfield  {author} {\bibinfo {author} {\bibfnamefont {Bernard}\ \bibnamefont
  {Carr}}\ and\ \bibinfo {author} {\bibfnamefont {Florian}\ \bibnamefont
  {Kuhnel}},\ }\bibfield  {title} {\enquote {\bibinfo {title} {Primordial black
  holes as dark matter: Recent developments},}\ }\href {\doibase
  10.1146/annurev-nucl-050520-125911} {\bibfield  {journal} {\bibinfo
  {journal} {Ann. Rev. Nucl. Part. Sci.}\ }\textbf {\bibinfo {volume} {70}},\
  \bibinfo {pages} {355--394} (\bibinfo {year} {2020})},\ \Eprint
  {http://arxiv.org/abs/2006.02838} {arXiv:2006.02838 [astro-ph.CO]}
  \BibitemShut {NoStop}%
\end{thebibliography}%
%%%%%%%%%%%%%%%%%%%%%%%%%%%%%%%%%%%%%%%%
%%%%%%%%%%%%%%%%%%%%%%%%%%%%%%%%%%%%%%%%

%%%%%%%%%%%%%%%%%%%%%%%%%%%%%%%%%%%%%%%%
%%%%%%%%%%%%%%%%%%%%%%%%%%%%%%%%%%%%%%%%
\end{document}